\documentclass[final,authoryear,5p,times,twocolumn]{elsarticle}
\usepackage{graphicx}
\usepackage{amssymb}
\usepackage{amsmath}
\usepackage{natbib}
\usepackage{threeparttable}
\journal{New Astronomy Reviews}

\newcommand{\beqa}{\begin{eqnarray}}
\newcommand{\eeqa}{\end{eqnarray}}
\newcommand{\be}{\begin{equation}}
\newcommand{\ee}{\end{equation}}
 \newcommand{\ba}{\begin{eqnarray}}
\newcommand{\ea}{\end{eqnarray}}
\newcommand{\omm}{{\Omega_{\rm M}}}

\begin{document}

\begin{frontmatter}

\title{Gamma-ray Burst Cosmology}

\author[ad1,ad2]{F. Y. Wang}
\address[ad1]{School of Astronomy and Space Science, Nanjing University, Nanjing 210093, China}
\address[ad2]{Modern Astronomy and Astrophysics (Nanjing University), Ministry of Education, Nanjing 210093, China}

\author[ad1,ad2]{Z. G. Dai}

\author[ad3]{E. W. Liang}
\address[ad3]{Department of Physics and GXU-NAOC Center for Astrophysics and Space Sciences, Guangxi University, Nanning 530004, China}

\begin{abstract}
Gamma-ray bursts (GRBs) are the most luminous electromagnetic
explosions in the Universe, which emit up to $8.8\times10^{54}$ erg
isotropic equivalent energy in the hard X-ray band. The high
luminosity makes them detectable out to the largest distances yet
explored in the Universe. GRBs, as bright beacons in the deep
Universe, would be the ideal tool to probe the properties of
high-redshift universe: including the cosmic expansion and dark
energy, star formation rate, the reionization epoch and the metal
enrichment history of the Universe. In this article, we review the
luminosity correlations of GRBs, and implications for constraining
the cosmological parameters and dark energy. Observations show that
the progenitors of long GRBs are massive stars. So it is expected
that long GRBs are tracers of star formation rate. We also review
the high-redshift star formation rate derived from GRBs, and
implications for the cosmic reionization history. The afterglows of
GRBs generally have broken power-law spectra, so it is possible to
extract intergalactic medium (IGM) absorption features. We also
present the capability of high-redshift GRBs to probe the
pre-galactic metal enrichment and the first stars.
\end{abstract}


\begin{keyword}

Gamma-ray bursts \sep cosmology \sep dark energy \sep star formation
rate \sep reionization.

\end{keyword}
\end{frontmatter}

\section{Introduction}
\label{sec:intro}

Gamma-ray bursts (GRBs) are among the most intriguing phenomena in
the Universe~\citep[for reviews,
see][]{Meszaros06,Zhang07,Gehrels09,Kumar14}. According to the
duration time $T_{90}$, GRBs are usually classified into two
classes: long GRBs ($T_{90}>2$ s) and short GRBs ($T_{90}<2$ s)
\citep{Kou93}. Long GRBs are thought to arise when a massive star
($\geq 25M_\odot$) undergoes core collapse, but the progenitors of
short GRBs are mergers of double neutron star or a neutron star and
a black hole binary \citep{Woosley06}. Long GRBs can be made
``relative standard candles", using luminosity correlations that
have been found in prompt and afterglow phases
\citep[i.e.,][]{Amati02,Ghirlanda04a,Liang05}. In view that the
history of the Universe during the so-called ``dark age" (from
cosmic background radiation at $z\sim 1100$ to the epoch when first
stars were formed around $z\sim 20$) is still poorly known
\citep{Barkana01}, GRBs, as bright beacons in the deep Universe,
would be the unique tool to illuminate the dark Universe and allow
us to unveil the reionization history. GRBs provide ideal probes of
the formation rate and environmental impact of stars in the
high-redshift universe, including the metal enrichment of the
intergalactic medium (IGM). Meanwhile, the infrared (IR) and near-IR
afterglows of long GRBs are also expected to be detectable out to
very high redshifts~\citep{Lamb00,Ciardi00,Bromm02,Wang12}. The
reason is that the cosmological time dilation translates higher
redshifts to earlier times in the source frame, at which the
afterglow is brighter. The afterglow intensities also decrease with
time. So according to the standard theory of GRBs, there is little
or no decrease in the flux of GRB afterglows at a given observed
time with increasing redshift for a single GRB. Consequently, GRBs
can be used as powerful probes of the very high redshift Universe.
The prospects of using GRBs as a cosmological tool are exciting:
\begin{itemize}
\item
The high luminosities of GRBs make them detectable out to high
redshifts.
\item
Gamma-ray photons suffer from no extinction when they propagate
towards us. But the optical photons from supernovae will suffer
extinction from the interstellar medium (ISM);
\item
The correlations between GRB spectral properties and energetics have
been shown to be powerful tools that ``standardize" GRB energetics.
So GRBs can be used to constrain cosmological parameters and the
nature of dark
energy~\citep[i.e.,][]{Dai04,Ghirlanda04b,Liang05,Ghirlanda06};
\item
The progenitors of GRBs are believed to be stellar mass objects. So
the intrinsic luminosity of GRB should not depend on the mass of
their host galaxy, which has small mass at high redshifts;
\item
Long GRBs triggered by the death of massive stars, which have been
shown to be associated with supernovae \citep{Stanek03,Hjorth03},
provide a complementary technique for measuring star formation rate.
They also represent a unique probe of the initial mass function and
the star formation of massive stars at very high redshifts. GRBs
offer the exciting opportunity to detect the Population~III (Pop
III) stars;
\item
GRB afterglows have smooth continuum spectra that allow one to
extract the IGM absorption features. The metal absorption lines in
the GRB spectra make GRBs powerful sources to study the metal
enrichment history;
\item The clean red damping wings of GRBs make them ideal tools to study the
reionization of IGM and ISM properties of their hosts.
\end{itemize}

First, GRBs can serves as the complementary tools to measure dark
energy and cosmic expansion. Type Ia supernovae (SNe Ia) are now
treated as ideal ``standard candles" for purposes of Hubble diagram.
In 1998, two teams studying distant SNe Ia discovered independent
evidence that the expansion of the Universe is speeding
up~\citep{Riess98,Perlmutter99}, which is attributed to the
mysterious component --- dark energy. The accelerated expansion of
the universe has also been confirmed by several independent
observations including those of the cosmic microwave background
(CMB)~\citep{Spergel03}, the baryonic acoustic oscillations
(BAO)~\citep{Eisenstein05}, X-ray gas mass fraction in galaxy
clusters \citep{Allen04}, and Hubble parameters
\citep{Jimenez03,Wangj14}. The standard $\Lambda$CDM model fits the
observational data well, but other dark energy models could not be
ruled out because of the precision of current data. The latest
observations confirm that about 70\% of the energy density of the
present Universe consists of dark energy. The direct evidence for
the current acceleration of the universe is related to the
observation of luminosity distances of high redshift SNe Ia. The SNe
Ia can be observed when accreting white dwarf stars exceed the mass
of the Chandrasekhar limit and explode. Thus they can be treated as
an ideal standard candle. But hitherto the highest redshift of SNe
Ia is $1.914$ \citep{Jones13}. GRBs are promising tools to study the
cosmic expansion at higher redshift, filling the gap between SNe Ia
and
CMB~\citep{Dai04,Ghirlanda04b,Liang05,Ghirlanda06,Wang06,Wang07,Schaefer07,Capozziello12,Wangf12}.
Similarly to SNe Ia, it has been proposed to use correlations of
GRBs between various properties of the prompt emission and of the
afterglow emission to standardize GRB
energetics~\citep[i,e.,][]{Ghirlanda04a,Xu05,Liang06,Firmani05,Amati06,Amati08,Liang08,Liang10}.

The high-redshift ($z>6$) star formation history (SFH) is important
in many fields in astrophysics. Direct star formation rate (SFR)
measurement beyond the reach of present instruments, particularly at
the low part of the galaxy luminosity function. Long GRBs triggered
by the death of massive stars, provide a complementary technique for
measuring the SFR \citep{Totani97,Wijers98,Bromm02}. Because the
lifetimes of massive stars are short, the SFR can be treated as
their death rate. Recent \emph{Swift} observation shows that GRBs
are not tracing the SFR exactly
\citep{Daigne06,Le07,Yuksel07,Salvaterra07,
Guetta07,Kistler08,Campisi10,Salvaterra12}. The SFR revealed by the
high-redshift GRBs seems to be much higher than that obtained from
high-redshift galaxy surveys \citep{Kistler08,Kistler09,Wang091}. An
enhancement about $(1+z)^{\delta}$ ($\delta\sim 0.5-1.5$) in the
observed rate of GRBs compared to SFR has been found
\citep{Kistler08,Robertson12,Wang13}. In order to explain this
discrepancy, many models have been proposed, including cosmic
metallicity evolution \citep{Li08,Qin10}, superconducting cosmic
strings \citep{Cheng10}, evolving star initial mass function
\citep{Wangf11,Xu08}, evolution of the luminosity function break of
GRBs \citep{Virgili11,Yu12}. But some other studies claimed that
there is no discrepancy between GRB rate and SFR
\citep{Elliott12,Hao13}. From host galaxy observations, long GRBs
prefer to form in a low-metallicity environment
\citep{Le03,Stanek06,Levesque14}, which is also required by
theoretical prediction. The mass loss of stars through winds is
proportional to the metallicity, so low-metallicity can reduce the
loss of angular momentum. Some studies have argued that GRB
progenitors must have a low metallicity
\citep{Woosley061,Meszaros06,Langer06}. Observations also show
differences in the population of GRB host galaxies compared to
expectations for an unbiased star-formation tracer
\citep{Tanvir04,Fruchter06,Svensson10}.

The metal enrichment history has several important consequences for
structure formation \citep{Madau01,Karlsson12}. An early phase of
metal injection may qualitatively change the character of star
formation, from a high-mass (Pop~III) mode to a normal, low-mass
dominated (Pop I/II) one, if the enrichment exceeds a `critical
metallicity' of $Z_{\rm crit}\sim 10^{-4} Z_{\odot}$
\citep{Bromm01,Schneider02,Schneider06}. This mode transition has
crucial implications, e.g., for the expected redshift distribution
of GRBs \citep{Bromm06,Campisi11,deSouza11}, for the cosmic
reionization \citep{Cen03,Wyithe03,Furlanetto05}, and for the
chemical abundance of low-metallicity stars
\citep{Qian01,Frebel07,Frebel09,Tumlinson10}. Absorption lines on
the spectra of bright background sources, such as GRBs or quasars,
are main sources of information about the chemical properties of
high-redshift Universe \citep{Oh02,Furlanetto03,Oppenheimer09}.
These lines are due to absorption by metals in low-ionization stages
which arise in the higher column-density gas associated with damped
Ly$\alpha$ absorbers (DLAs). GRBs as bright sources have a number of
advantages compared to traditional lighthouses such as quasars
\citep{Bromm12}. Their number density drops much less precipitously
than quasars at $z>6$ \citep{Fan06}, together with the power-law
character of their spectra, renders them ideal probes of the early
IGM. Quasars show strong spectral features, such as broad emission
lines or the so-called ``blue bump" that complicate the extraction
of IGM absorption features. The analysis of the spectrum of distant
GRB~050904 has offered a wealth of detailed insight into the
physical conditions of the host galaxy at $z\simeq 6.3$
\citep{Totani06}. Salvaterra et al. (2009) claimed that they
identified two absorption lines (Si IV and Fe II) in the spectrum of
GRB~090423, the most distant spectroscopically confirmed burst at
$z=8.2$ \citep{Salvaterra09,Tanvir09}.

Besides the above fields, high-redshift GRBs also may be useful tools to
study dark matter and primordial non-Gaussianities. Because the
matter power spectrum is dependent on the mass of dark matter, i.e.,
cold, warm and hot, which can result in different structure formation.
So the number counts of high-redshift GRBs can set strong lower
limits on the dark matter mass \citep{Mesinger05,deSouza13}. The
rate of high-redshift GRBs is also dependent on the amount of
primordial non-Gaussianity in the density field \citep{Maio12}.

In this work, we review the cosmological implications of GRBs. Then
in the following sections the gamma-ray bursts cosmology is
reviewed: The second section is dedicated to luminosity correlations
and cosmological constraints from GRBs. Section 3 discusses the
capability of GRBs to reveal the high-redshift SFR and cosmic
reionization. In section 4, the capability of GRBs to probe the
metal enrichment history is discussed. The last section provides a
summary and future prospect.

\section{GRBs as standard candles to probe dark energy}
The best way to measure properties of the dark energy is to measure
the expansion history of our universe, i.e., the redshift-distance
relation. To this end, WFIRST has been proposed to determine the
distances of 1000 SNe Ia with exquisite accuracy. To explore the
properties of dark energy, the best method is to measure it over a
wide range of redshifts, but SNe Ia can only be detected at low
redshifts, i.e., $z<2.0$. {GRBs can extend the Hubble diagram to
high redshifts. So many attempts have been performed to standardize
GRBs. \cite{Frail01} found that the collimated energetics of GRBs
clustered around $5\times 10^{50}$ erg, which was confirmed by
\cite{Bloom03}. Observations also require that the GRB emission is
only in a small angle \citep{Waxman98,Fruchter99}. The collimated
GRB model predicts that the appearance of an achromatic break in the
afterglow light curve \citep{Rhoads97,Sari99}. This break is
important to standardize the energetic of GRBs.

Similar as SNe Ia, the luminosity correlations are required to probe
dark energy using GRBs. In this section, we first review luminosity
correlations of GRBs. Then the progress on dark energy revealed by
GRBs is discussed. Some reviews have discussed this topic
\citep[i.e.,][]{Ghirlanda06,Dai07,Capozziello12,Amati13}.

\subsection{The luminosity correlations of GRBs}
The luminosity correlations are connections between parameters of
the light curves and/or spectra with the GRB luminosity or energy.
The isotropic luminosity can be calculated as \be
 L = 4\pi d^2_{L}P_{\rm bolo} \; \label{ldl}
\ee and the total collimation-corrected energy is  \be
E_{\gamma}=E_{\rm iso}F_{\rm beam}=4\pi d^2_{L}S_{\rm bolo}F_{\rm
beam}(1+z)^{-1}. \; \label{egdl} \ee Here, $P_{\rm bolo}$ and
$S_{\rm bolo}$ are the bolometric peak flux and fluence,
respectively, while $F_{\rm beam} = 1 - \cos{\theta_{\rm jet}}$ is
the beaming factor with jet opening angle $\theta_{\rm jet}$. The
peak fluxes and fluences are given over a wide variety of observed
bandpasses, and with the wide range of redshifts which correspond to
different range of energy bands in the rest frame of GRB. So the
K-correction is important \citep{Bloom01}. $P_{\rm bolo}$ and
$S_{\rm bolo}$ are computed from the differential energy spectrum
$\Phi(E)$ as follows:

\begin{equation}
P_{\rm bolo} = P  \ {\times} \ \frac{\int_{1/(1 + z)}^{10^4/(1 +
z)}{E \Phi(E) dE}} {\int_{E_{\rm min}}^{E_{\rm max}}{\Phi(E) dE}} \
, \label{eq: defpbolo}
\end{equation}

\begin{equation}
S_{\rm bolo} = S \ {\times} \ \frac{\int_{1/(1 + z)}^{10^4/(1 +
z)}{E \Phi(E) dE}} {\int_{E_{\rm min}}^{E_{\rm max}}{E \Phi(E) dE}}
\ , \label{eq: defsbolo}
\end{equation}
with $P$ and $S$ being the observed peak energy and fluence in units
of ${\rm photons/cm^2/s}$ and ${\rm erg/cm^2}$, respectively, and
$(E_{\rm min}, E_{\rm max})$ the detection thresholds of the
observing instrument. In general, the differential energy spectrum
is modeled using a broken power\,-\,law \citep{Band93},
\begin{equation}
\Phi(E) = \left \{
\begin{array}{ll}
A E^{\alpha} {\rm e}^{-(2 + \alpha) E/E_{\rm peak}} & E \le
\frac{\alpha
-\beta}{2 + \alpha}E_{\rm peak} \\ ~ & ~ \\
B E^{\beta} & {\rm otherwise}
\end{array}
\right . \  \label{eq: band}
\end{equation}
where $\alpha$ is the power-law index for photon energies below the
break and $\beta$ is the power-law index for photon energies above
the break. Some differential energy spectra of GRBs also show
power-law spectra plus an exponential cutoff. The luminosity
distance $d_L$ can be expressed as
\begin{eqnarray}
 d_L(z) = (1+z)\frac{c}{H_0} \int_0^z \frac{dz'}{E(z')} \; ,
 \label{dlum1}
\end{eqnarray}
where $E^2(z) = \omm (1+z)^3 + \Omega_{\rm x}f_{\rm x}(z)$ and
$f_{\rm x}(z)$ is given by
\begin{eqnarray}
  \label{eq:fz}
  f_{\rm x}(z)=\exp \left[
    3\int_0^z\frac{1+w(\tilde{z})}{1+\tilde{z}}\mathrm{d}\tilde{z}
  \right]
  ,
\end{eqnarray}
where $w(z)$ is the equation of state (EOS) of dark energy. For
$\Lambda$CDM, Eq.(\ref{dlum1}) reduces to
\begin{eqnarray}
 d_L(z) = (1+z)\frac{c}{H_0} \int_0^z \frac{dz'}{\Omega_M(1+z')^3)+\Omega_\Lambda} \;
 .
\end{eqnarray}

\begin{figure}
\includegraphics[angle=0,width=0.5\textwidth]{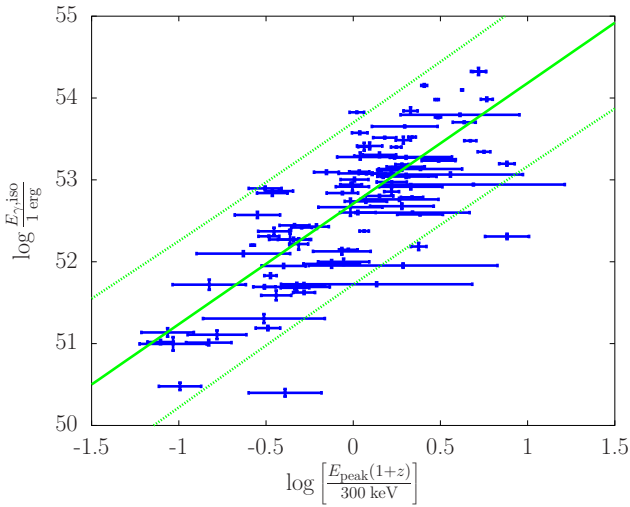}
\caption{The Amati correlation of 101 GRBs. Solid line is the best
fit, and dashed lines represent the $1\sigma$ dispersion. (Adapted
from Figure 1 in \cite{Wang11}.) }\label{amatirelation}
\end{figure}

The promising luminosity correlations of GRBs are as follows:
\begin{itemize}
\item
$L_{\rm iso}-\tau_{\rm lag}$ correlation. The luminosity-time lag
correlation was first discovered by \cite{Norris00} based on six
GRBs observed by BATSE with optical redshifts \citep[also
see][]{Schaefer01}, which was confirmed by GRBs observed by
\emph{Swift} \citep{Gehrels06} including the peculiar long GRB
060218 \citep{Liang06}. But this correlation is challenged by recent
study \citep{Bernardini15}. This correlation shows that more
luminous bursts are also characterized by shorter time lags, i.e.,
$L_{\rm iso}\propto \tau_{\rm lag}^{-1.25}$ \citep{Norris00}. This
correlation has been used as a redshift indicator to estimate $z$
for GRBs \citep{Band04}, and to constrain cosmological parameters
and dark energy \citep{Schaefer07,Wang07}. The interpretation of
this correlation is various, including viewing angle of the
collimated jet \citep{Ioka01} or radiative cooling effect
\citep{Crider99}.
\item
$L_{\rm iso}-V$ correlation. Fenimore and Ramirez-Ruiz (2000) found
that the time variability is correlated with the luminosity of GRBs,
which indicates that more luminous bursts have more variable light
curves. Later, this correlation is confirmed with more GRBs by
\cite{Reichart01} and \cite{Schaefer01}. But the intrinsic scatter
of $L_{\rm iso}-V$ is very large \citep{Guidorzi06,Wang11}, and the
index is still under debate \citep{Li06}. The variability $V$ is
different for various instruments. The origin of the $L_{\rm iso}-V$
correlation may be based on the screening effect of the photosphere
\citep{Kobayashi02,Meszaros02}.
\item
Amati correlation. Amati et al (2002) found that the isotropic
energy $E_{\rm iso}$ is correlated with the rest-frame peak energy
of the prompt spectrum, i.e., $E_{\rm peak}\propto E_{\rm
iso}^{0.52}$. Subsequent observations with various detectors (such
as \emph{Swift} and \emph{Fermi}) confirmed this correlation
\citep{Amati09,Ghirlanda10,Sakamoto11,Wang11}. Moreover, it was
found that the Amati correlation also holds within individual GRBs
using time-resolved spectra, and the slopes are consistent with the
correlation from time-integrated spectra
\citep{Ghirlanda10,Frontera12}. The possible interpretations of
Amati correlation include the synchrotron mechanism in relativistic
shocks \citep{Zhang02} and emission from off-axis relativistic jets
\citep{Yamazaki04,Eichler04}. Figure \ref{amatirelation} shows the
Amati correlation including 101 GRBs. The intrinsic scatter is 0.62.
\item
Yonetoku correlation. The correlation $L_{\rm iso}\propto E_{\rm
peak}^2$ was found with a sample of 16 GRBs
\citep{Yonetoku04,Wei03}. This correlation was confirmed by
\cite{Liang04}. Similar as Amati correlation, $E_{\rm peak}-L_{\rm
iso}$ also holds within individual GRBs using time-resolved spectra
\citep{Ghirlanda10}. The possible origin of this correlation is
similar as that of Amati correlation. Figure \ref{Yonetoku} shows
the Yonetoku correlation. The intrinsic scatter of this correlation
is 0.62 \citep{Wang11}.
\item
Ghirlanda correlation. A tight correlation between spectral peak
energy $E_{\rm peak}$ and collimated energy $E_\gamma$ was
discovered by \cite{Ghirlanda04a} using 15 GRBs. The intrinsic
scatter is up to $0.1$, so this correlation is a promising tool to
constrain dark energy \citep{Dai04,Ghirlanda04b}. By considering the
wind circumburst density, \cite{Nava06} found this correlation also
holds, and the intrinsic scatter is even smaller. One of the major
challenge for this correlation is that most of GRBs observed by
\emph{Swift} do not show achromatic breaks in the afterglow light
curve \citep{Willingale07}. So the break time is very hard to
determine. This correlation can be understood within the annular jet
model \citep{Eichler06} and photosphere model \citep{Thompson06}.
The latest Ghirlanda correlation is shown in Figure \ref{Egamma}.
\item
$L_{\rm iso}-E_{\rm peak}-T_{0.45}$ correlation. \cite{Firmani06}
found a tight correlation $L_{\rm iso}\propto E_{\rm
peak}^{1.62}T_{0.45}^{-0.49}$ using parameters in prompt emission
only. $T_{0.45}$ represents the ``high signal" timescale.
\item
Liang-Zhang correlation. Without imposing any theoretical model,
\cite{Liang05} found an empirical correlation among the isotropic
energy of the prompt gamma-ray emission $E_{\rm iso}$, the
rest-frame peak energy $E_{\rm peak}$, the rest-frame break time in
the optical band $t_{\rm break}$ using 15 bursts. The correlation
reads $E_{\rm peak}\propto E_{\rm iso}^{0.52}t_{\rm break}^{0.64}$.
If we take the optical break time as the jet break time, this
correlation is similar to Ghirlanda correlation. The intrinsic
scatter of this correlation is also very small, so it could be used
to constrain dark energy \citep{Liang05,Wang06,Wei13}. The
Liang-Zhang correlation is shown in Figure \ref{LiangZhang}.
\item
$L_X-T_a$ correlation. \cite{Dainotti08} discovered a tight
correlation between X-ray luminosity $L_X$ and $T_a$, where $T_a$ is
the time at which the X-ray light curve establishes a afterglow
power-law decay \citep{Willingale07}. The intrinsic scatter of this
correlation is about 0.33. By adding a third parameter isotropic
energy $E_{\rm iso}$, \cite{Xu12} found a new correlation, i.e.,
$L_X\propto T_a^{-0.87}E_{\rm iso}^{0.88}$.
\end{itemize}

\begin{figure}
\includegraphics[angle=0,width=0.5\textwidth]{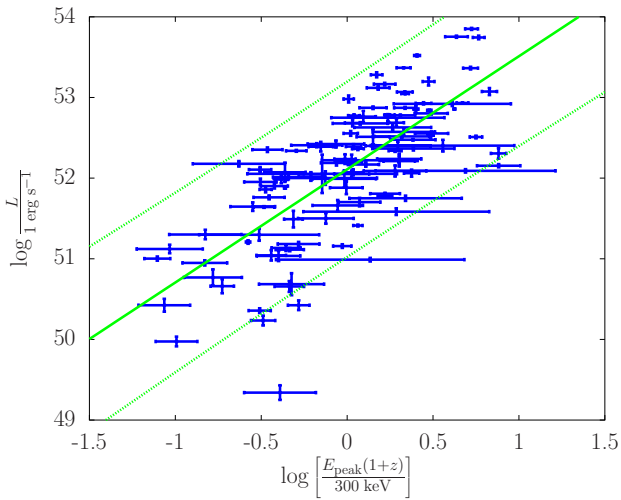}
\caption{The Yonetoku correlation. Solid line is the best fit, and
dashed lines represent the $1\sigma$ dispersion. (Adapted from
Figure 1 in \cite{Wang11}.)}\label{Yonetoku}
\end{figure}

\begin{figure}
\includegraphics[angle=0,width=0.5\textwidth]{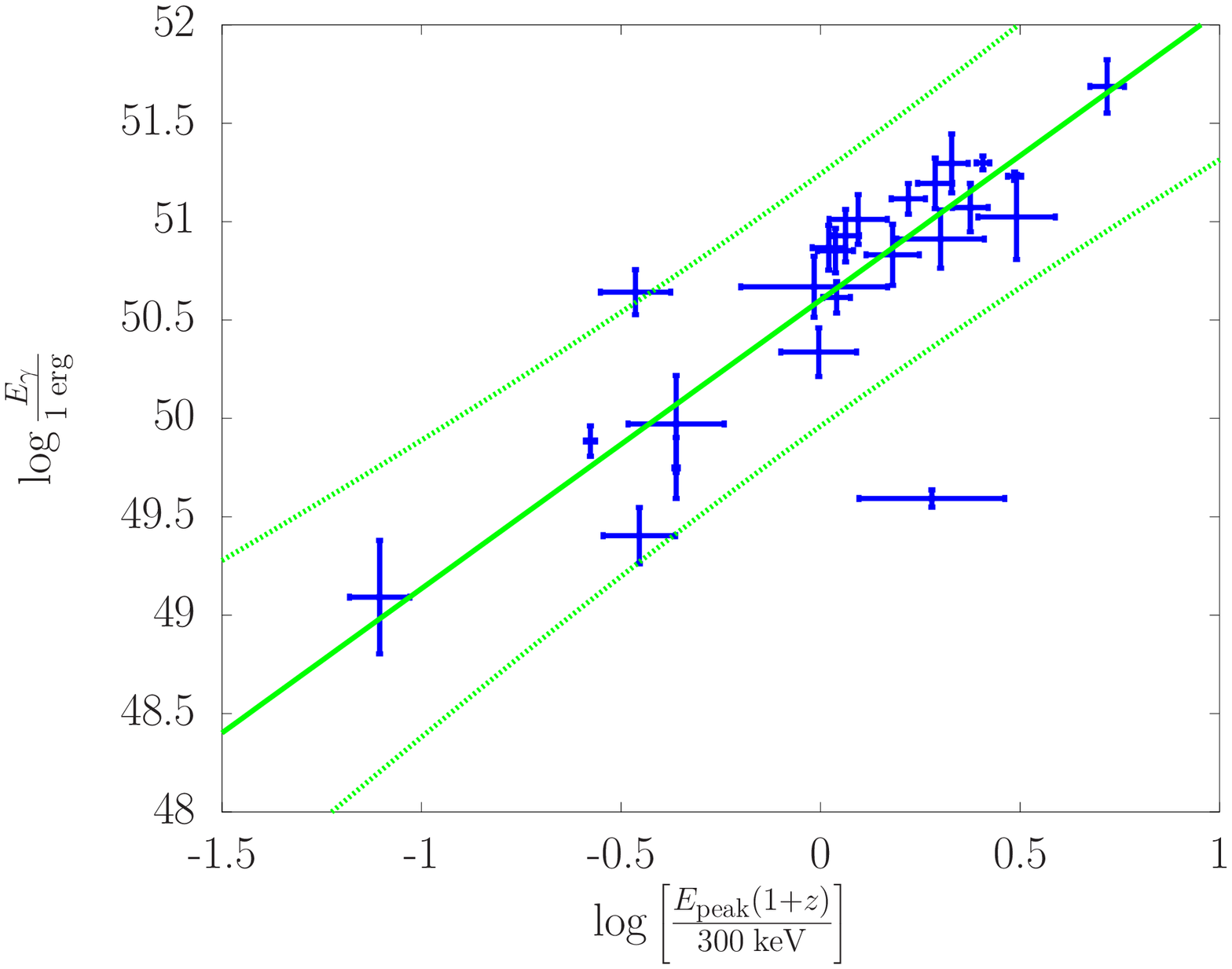}
\caption{The Ghirlanda correlation. Solid line is the best fit, and
dashed lines represent the $1\sigma$ dispersion. (Adapted from
Figure 1 in \cite{Wang11}.)}\label{Egamma}
\end{figure}

\begin{figure}
\includegraphics[angle=0,width=0.5\textwidth]{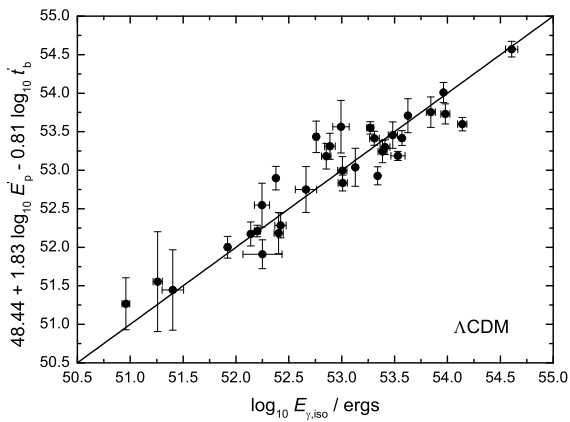}
\caption{The Liang-Zhang correlation of 33 GRBs in the $\Lambda$CDM
model. (Adapted from Figure 1 in \cite{Wei13}.)}\label{LiangZhang}
\end{figure}

Due to GRBs cover large redshift range, whether the correlations
evolve with the redshift should be discussed. It's found that the
slope of Amati correlation may vary with redshift significantly
using a small sample of GRBs \citep{Li07}. \cite{Basilakos08} found
no statistically significant evidence for redshift dependence of
slopes in five correlations using 69 GRBs compiled by
\cite{Schaefer07}. \cite{Wang11} enlarge the GRB sample and test six
GRB correlations. There is no statistically significant evidence for
the evolution of the luminosity correlations with redshift is found.
The slopes of correlations versus redshift are all consistent with
zero at the 2$\sigma$ confidence level.

The instrumental selection effects may affect the observed
luminosity correlations. Some outliers to these correlations have
been discovered (Soderberg et al. 2004; Vaughan et al. 2006; Campana
et al. 2007; Rizzuto et al. 2007; Urata et al. 2009). Nakar and
Piran (2005) considered the different samples of GRBs detected by
BATSE. They found that a large fraction of GRBs were inconsistent
with the Amati correlation by assuming all possible redshifts. Band
and Preece (2005) also found that about 88 and 1.6 per cent of their
BATSE GRBs were outliers to the Amati and Ghirlanda correlations,
respectively. Butler et al. (2007) claimed that the Amati
correlation exists, but it may be due to the selection effect. Using
a large sample of GRBs with pseudo-redshifts determined by the
$L_{\rm iso}-\tau_{\rm lag}$ correlation, Ghirlanda et al. (2005)
argued that the Amati correlation really existed by taking into
account the intrinsic scatter. By considering the triggering
threshold limits for several GRB detectors, Nava et al. (2008) and
Ghirlanda et al. (2008) claimed that only 6\% of BATSE long GRBs are
inconsistent with Amati correlation, while on outliers to the
Ghirlanda correlation are found. By simulating the BATSE Large Area
Detectors¡¯ triggering thresholds, Shahmoradi and Nemiroff (2011)
found that the Amati and Ghirlanda correlations are statistically
real but strongly affected by the thresholds of GRB detectors.
Ghirlanda et al. (2012) studied the selection effect on the Yonetoku
correlation using Monte Carlo simulations, and found this
correlation must be physical. Dainotti et al. (2014) proposed a
general method to check the selection effects for GRB correlations
and found the $L_x-T_a$ correlation is not generated by the biases.
Interestingly, using the time-resolved spectra, similar correlations
were found in individual bursts (Firmani et al. 2009; Ghirlanda et
al. 2010; 2011). This strongly supports that the correlations are
physical.

\subsection{Constraints on dark energy and cosmological parameters}
The most common method to constrain dark energy is through its
influence on the expansion history of the universe, which can be
extracted from the luminosity distance $d_L(z)$ and the angular
diameter distance $d_A(z)$. In addition, the weak gravitational
lensing, growth of large-scale structure, and redshift space
distortion can also provide useful constraints on dark energy.
Theoretical models can be tested using the $\chi^2$ statistic. The
typical way to probe dark energy from standard candles is as
follows. With luminosity distance $d_{L}$ in units of megaparsecs,
the theoretically predicted distance modulus is
\begin{equation}
\mu=5\log(d_{L})+25.
\end{equation}
The likelihood functions for the cosmological parameters can be
determined from $\chi^{2}$ statistics,
\begin{equation}
\chi^{2}(\Omega_{M},\Omega_{DE})=\sum_{i=1}^{N}
\frac{[\mu_{i}(z_{i},H_{0},\Omega_{M},\Omega_{DE})-\mu_{0,i}]^{2}}
{\sigma_{\mu_{0,i}}^{2}},
\end{equation}
where $\mu_{0,i}$ is the observed distance modulus, and
$\sigma_{\mu_{0,i}}$ is the standard deviation in the individual
distance modulus. The confidence regions in the
$\Omega_{M}-\Omega_{DE}$ plane can be found through marginalizing
the likelihood functions over $H_{0}$ (i.e., integrating the
probability density $p\propto\exp(-\chi^2/2)$ for all values of
$H_{0}$).

A lot of effort had been made to constrain cosmological parameters
using GRBs since their cosmological origin was confirmed.
\cite{Schaefer03} obtained the first GRB Hubble diagram based on
$L_{\rm iso}-V$ correlation, and found the mass density
$\Omega_M<0.35$ at the $1\sigma$ confidence level. After
\cite{Ghirlanda04a} found the Ghirlanda correlation, \cite{Dai04}
first used this correlation with 12 bursts and found the mass
density $\Omega_M=0.35\pm0.15$ at the $1\sigma$ confident level for
a flat universe by assuming that some physical explanation comes
into existence. \cite{Ghirlanda04b} using 14 GRBs and SNe Ia
obtained $\Omega_{\rm M}=0.37\pm0.10$ and $\Omega_{\Lambda}=0.87\pm
0.23$. Assuming a flat universe, the cosmological parameters were
constrained to be $\Omega_{\rm M}=0.29\pm0.04$ and
$\Omega_{\Lambda}=0.71\pm 0.05$ \citep{Ghirlanda04b}. For a flat
universe, \cite{Firmani05} found $\Omega_{M}=0.28\pm0.03$ and
$z_{T}=0.73\pm0.09$ for the combined GRB and SNe Ia sample, where
$z_{T}$ is the transition redshift, at which the universe starts
accelerating expansion. In the dark energy model of $w_{z}=w_{0}$,
they found $\Omega_{M}=0.44$ and $w_{0}=-1.68$ with $z_{T}=0.40$ for
the combined GRB and SNe Ia sample. \cite{Xu05} obtained
$\Omega_{M}=0.15^{+0.45}_{-0.13}(1\sigma)$ using 17 GRBs. Friedman
\& Bloom (2005) discussed several possible sources of systematic
errors in using GRBs as standard candles. Using the $E_{\rm
peak}-E_{\rm iso}-t_{\rm break}$ correlation, \cite{Liang05} found
the $1\sigma$ constraints are $0.13<\Omega_{\rm{M}}<0.49$ and
$0.50<\Omega_{\Lambda}<0.85$ for a flat universe. They also found
the transition redshift is $0.78^{+0.32}_{-0.23}$. \cite{Wang06}
using the Liang-Zhang correlation to investigate the transition
redshifts in different dark energy models via GRBs and SNe Ia, see
also \cite{Wangf06}. \cite{DiGirolamo05} simulated different samples
of gamma-ray bursts and found that $\Omega_M$ could be determined
with accuracy $\sim$ 7\% with data from 300 GRBs. Meanwhile, many
works have been done on this field, such as \cite{Mortsell05},
\cite{Bertolami06}, \cite{Firmani06}, \cite{Hooper07}, \cite{Li08a},
\cite{Mosquera08}, \cite{Basilakos08}, \cite{Wang08}, \cite{Wang09},
\cite{Yu09}, \cite{Cardone09}, \cite{Qi10}, \cite{Cardone11},
\cite{Demianski11a}, \cite{Demianski11b}, and \cite{Pan13}. More
recently, theoretical arguments and observational evidence both
suggested that a small fraction of fast radio bursts (FRBs) may be
associated with GRBs \citep{Bannister12,Thornton13,Zhang14}. So the
dispersion measure from FRBs and redshifts from GRBs makes these
systems a plausible tool to study cosmological parameters
\citep{Deng14,Zhou14,Gao14}.

Unfortunately, because of lack of low-$z$ GRBs, the luminosity
correlations has been obtained only from moderate-$z$ GRBs. So this
correlation is cosmology-dependent, i.e., the isotropic energy
$E_{\rm iso}$, collimation-corrected energy $E_\gamma$, and
luminosity $L_{\rm iso}$ are as functions of luminosity distance
$d_L$, which is dependent on cosmology model. This is the so-called
``circularity problem" of GRBs. In the following, we present
different methods to overcome this problem.

The first method is fitting the cosmological parameters and
luminosity correlation simultaneously. \cite{Ghirlanda04b} first
used this method to overcome the ``circularity problem". Schaefer
(2007) used 69 GRBs and five correlations to constrain cosmological
parameters. \cite{Wang07} also used 69 GRBs and other cosmological
probes to constrain cosmological parameters. They make simultaneous
uses of five luminosity indicators, which are correlations of
$L_{\rm iso}-\tau_{\rm lag}$, $L_{\rm iso}-V$, $E_{\rm peak}-L_{\rm
iso}$, $E_{\rm peak}-E_{\gamma}$, and $\tau_{\rm RT}-L_{\rm iso}$.
After obtaining the distance modulus of each burst using one of
these correlations, the real distance modulus can be calculated,
\begin{equation}
\mu_{\rm fit}=(\sum_i \mu_i/\sigma_{\mu_i}^2)/(\sum_i
\sigma_{\mu_i}^{-2}),
\end{equation}
where the summation runs from $1-5$ over the correlations with
available data, $\mu_i$ is the best estimated distance modulus from
the $i$-th relation, and $\sigma_{\mu_i}$ is the corresponding
uncertainty. The uncertainty of the distance modulus for each burst
is
\begin{equation}
\sigma_{\mu_{\rm fit}}=(\sum_i \sigma_{\mu_i}^{-2})^{-1/2}.
\end{equation}
Because each correlation with its limitations and possible biases,
simultaneous using may affect the cosmological indicators. But in
order to enlarge the sample, some attempts have been performed. When
calculating constraints on cosmological parameters and dark energy,
the normalizations and slopes of the five correlations are
marginalized. The marginalization method is to integrate over some
parameters for all of its possible values. The $\chi^2$ value is
\begin{equation}
\chi^{2}_{\rm GRB}=\sum_{i=1}^{N}
\frac{[\mu_{i}(z_{i},H_{0},\Omega_{M},\Omega_{DE})-\mu_{{\rm
fit},i}]^{2}}{\sigma_{\mu_{{\rm fit},i}}^{2}},
\end{equation}
where $\mu_{{\rm fit},i}$ and $\sigma_{\mu_{{\rm fit},i}}$ are the
fitted distance modulus and its error. In addition to GRBs, SNe Ia,
CMB, BAO, X-ray gas mass fraction in galaxy clusters and growth rate
data are also ideal cosmological probes. CMB is the remnant of the
cosmic recombination epoch. It contains abundant information of the
early universe. The positions of the acoustic peaks contains the
information of dark energy \citep{Peebles70,Bond84}. The shift
parameter is defined as \citep{Bond97,WangY06}
\begin{equation}
\label{shift}
\mathcal{R}=\frac{\sqrt{\Omega_M}}{\sqrt{|\Omega_k|}}{\rm
sinn}\left(\sqrt{|\Omega_k|}\int_{0}^{z_{\rm
ls}}\frac{dz}{E(z)}\right)=1.70\pm 0.03,
\end{equation}
where $E(z) \equiv H(z)/H_0$ and the function ${\rm sinn}(x)$ is
defined as ${\rm sinn}(x) = \sin(x)$ for a closed universe, ${\rm
sinn}(x) = \sinh(x)$ for an open universe and ${\rm sinn}(x) = x$
for a flat universe. The last scattering redshift $z_{\rm ls}$ can
be fitted as
\begin{equation}
z_{\rm ls}=1048[1+0.00124 (\Omega_{b}h^2)^{-0.738}][1+g_1
(\Omega_{M}h^2)^{g_2}],
\end{equation}
where the quantities $g_1$ and $g_2$ are defined as $
g_1=0.078(\Omega_b h^2)^{-0.238} [1+39.5 (\Omega_b
h^2)^{0.763}]^{-1}$ and $g_2=0.56 [1+21.1 (\Omega_b
h^2)^{1.81}]^{-1}$ respectively \citep{Hu96}. The $\chi^2$ value is
\begin{equation}
\chi^{2}_{\rm CMB}=\frac{(\mathcal{R}-1.70)^2}{0.03^2}.
\end{equation}

BAO refers to an overdensity of baryonic matter due to acoustic
waves which propagated in the early universe
\citep{Silk68,Peebles70}. BAO provides a ``standard ruler" for
length scale in cosmology to explore the expansion history of the
universe. The acoustic signatures in the large-scale clustering of
galaxies can be used to constrain cosmological parameters by
detection of a peak in the correlation function
\citep{Eisenstein05}. The $A$ parameter is defined as
\begin{equation}
A = \frac{\sqrt{\Omega_{M}}}{z_1}
 \left[\frac{z_1}{E(z_1)}\frac{1}{|\Omega_k|} {\rm sinn}^2
 \left(\sqrt{|\Omega_k|}\int_0^{z_1}\frac{dz}{E(z)}\right)\right]^{1/3},
\end{equation}
measured from the SDSS data to be $A=0.469(0.95/0.98)^{-0.35}\pm
0.017$, where $z_1 = 0.35$. The $\chi^2$ value is
\begin{equation}
\chi^{2}_{\rm BAO}=\frac{(A-0.469)^2}{0.017^2}.
\end{equation}

\begin{figure}
\includegraphics[angle=0, width=0.5\textwidth]{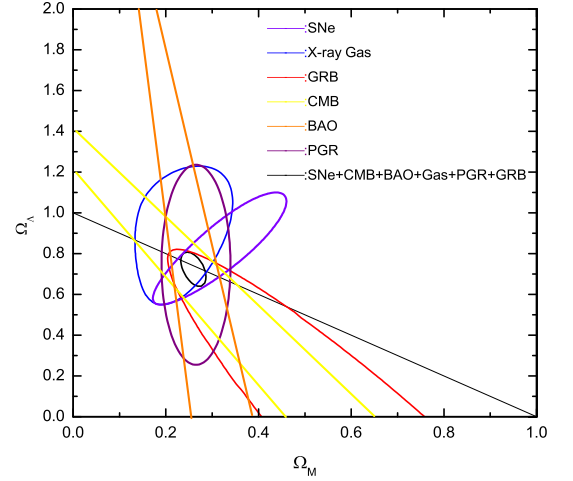} \caption{Joint confidence intervals of $1\sigma$
for $(\Omega _M ,\Omega _\Lambda)$ from the observational datasets.
The thick black line contour, the blue contour, the red contour, the
yellow contour, the violet contour, the orange contour, and the
purple contour show constraints from from all the datasets, 26
galaxy clusters, 69 GRBs, CMB shift parameter, 182 SNe Ia, BAO, and
2dF Galaxy Redshift Survey, respectively. The thin solid line
represents a flat universe. (Adapted from Figure 2 in
\cite{Wang07}.) \label{wangF07}}
\end{figure}

In Figure \ref{wangF07}, the constraint on the $\Lambda$CDM model is
shown. Different color contours represent constraints from different
data. The best fitted parameters are consistent with a flat
geometry. Figure \ref{wang071} shows the cosmological constrain on
the $w=w_0$ model from SNe Ia (blue) and GRBs (red). The combined
constraint is shown as solid contours. The constraint is much
tighter by adding the GRB data. \cite{Li08b} also performed a Markov
Chain Monte Carlo (MCMC) global fitting analysis to overcome the
circularity problem. The Ghirlanda correlation and 27 GRBs are used.
They treated the slopes of Ghirlanda correlation and cosmological
parameters as free parameters and determine them simultaneously
through MCMC analysis on GRB data together with other observational
data, such as SNe Ia, CMB and large-scale structure (LSS).
\cite{Amati08} measured the cosmological parameters using Amati
correlation using global fitting method. The extrinsic scatter was
assumed on the parameter of $E_{\rm peak}$, but the
cosmological-dependent value is $E_{\rm iso}$ \citep{Ghirlanda09}.

\begin{figure}
\includegraphics[angle=0, width=0.5\textwidth]{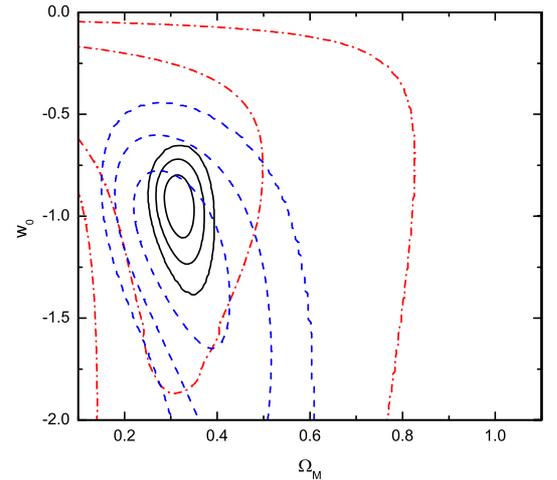} \caption{Joint confidence intervals of $1\sigma$
to $3\sigma$ for $(\Omega _M ,w_{0})$ from the observational
datasets. The solid contours, the dashed blue contours, and the
dot-dashed red contours show constraints from all the datasets, 26
galaxy clusters, and 69 GRBs, respectively. (Adapted from Figure 4
in \cite{Wang07}.) \label{wang071}}
\end{figure}

The second method is to calibrate the correlations of GRBs using SNe
Ia data at low redshifts. The principle of this method is that
objects at the same redshift should have the same luminosity
distance in any cosmology model. Therefore, the luminosity distance
at any redshift in the redshift range of GRBs can be obtained by
interpolating (or by other approaches) directly from the SNe Ia
Hubble diagram. Then if further assuming these calibrated GRB
correlations to be valid for all long GRB data, the standard Hubble
diagram method to constrain the cosmological parameters from the GRB
data at high redshifts obtained by utilizing the correlations.
\cite{Liang08} first calibrated the GRB correlations using
an interpolation method. The error of distance modulus of linear
interpolation can be calculated as \citep{Liang08,Wei10}
\begin{equation}
\sigma_{\mu}=([(z_{i+1}-z)/(z_{i+1}-z_i)]^2\epsilon_{\mu,i}^2+[(z-z_{i})/(z_{i+1}-z_i)]^2\epsilon_{\mu,i+1}^2)^{1/2},
\end{equation}
where $\epsilon_{\mu,i}$ and $\epsilon_{\mu,i+1}$ are errors of the
SNe Ia, $\mu_{i}$ and  $\mu_{i+1}$ are the distance moduli of the
SNe Ia at $z_{i}$ and $z_{i+1}$ respectively. Similar to the
interpolation method, \cite{Cardone09} constructed an updated GRBs
Hubble diagram on six correlations calibrated by local regression
from SNe Ia. \cite{Kodama08} presented that the $L_{\rm iso}$ -
$E_{\rm peak}$ correlation can be calibrated with the empirical
formula fitted from the luminosity distance of SNe Ia. This method
has been used to constrain cosmological parameters by combining
these GRB data with SNe Ia in a following work by \cite{Tsutsui08}.

However, it must be noted that this calibration procedure depends
seriously on the choice of the formula and various possible formulas
can be fitted from the SNe Ia data that could give different
calibration results of GRBs. As the cosmological constraints from
GRBs are sensitive to GRBs calibration results \citep{Wang08}, the
reliability of this method should be tested carefully. Moreover, as
pointed out by \cite{Wang08}, the GRB luminosity correlations which
are calibrated by this way are no longer completely independent of
all the SNe Ia data points. Therefore these GRB data can not be used
to directly combine with the whole SNe Ia dataset to constrain
cosmological parameters and dark energy. In order to search a unique
expression of the fitting formula, \cite{Wangf09} used the
cosmographic parameters
\citep{Capozziello08,Vitagliano10,Xia12,Gao12}. The luminosity
distance can be expanded as \citep{Visser04} \beqa
d_L={{c}\over{H_0}}\left\{z+{{1\over2}(1-q_0)}z^2-{{1}\over{6}}
\left(1-q_0-3q_0^2+j_0\right)z^3\right.
\nonumber\\
+{{1}\over{24}}\left[2-2q_0-15q_{0}^{2}-15q_0^3+5j_0 +10q_0
j_0+s_0\right]z^4 +O(z^5)\},\label{cosmgra} \eeqa where $q$ is the
deceleration parameter, $j$ is the so-called ``jerk'', and $s$ is
the so-called ``snap'' parameter. These quantities are defined as
\begin{equation}
q=-\frac{1}{H^2}\frac{\ddot{a}}{a};
\end{equation}
\begin{equation}
j=\frac{1}{H^3}\frac{\dot{\ddot{a}}}{a};
\end{equation}
\begin{equation}
s=\frac{1}{H^4}\frac{\ddot{\ddot{a}}}{a}.
\end{equation}
Equation (\ref{cosmgra}) is only dependent on the cosmological
principles and FRW metric, so the expansion is model-independent. But
the Taylor-expansion of $d_L$ is not valid at $z>1$. So expansion
$d_L$ as a function of $y=z/1+z$ is much useful
\citep{Cattoen07,Wangfy11}. The calibrated Hubble diagram of GRBs is
shown in Figure \ref{hubble} \citep{Wangfy11}.

\begin{figure}
\includegraphics[angle=0, width=0.5\textwidth]{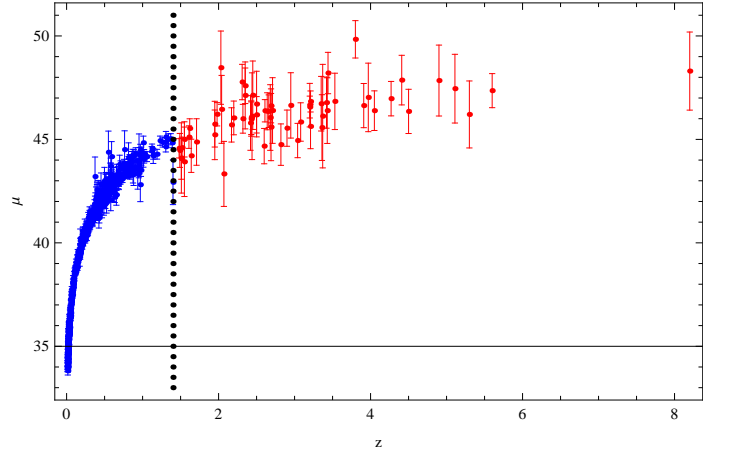} \caption{
The Hubble diagram of 557 SNe Ia (blue) and 66 high-redshift GRBs
(red). (Adapted from Figure 2 in \cite{Wangfy11}.) \label{hubble}}
\end{figure}

The third method is to calibrate the standard candles using GRBs in
a narrow redshift range ($\delta z$) near a fiducial redshift
\citep{Lamb05,Liang06,Ghirlanda06b}. \cite{Liang06} proposed a
procedure to calibrate the Liang-Zhang correlation with a sample of
GRBs in a narrow redshift range. No low-redshift GRB sample is
needed in this method. The calibration procedure can be described as
follows. First, calibrate the power-law index of Liang-Zhang
correlation using a sample of GRBs that satisfy this correlation and
are distributed in a narrow redshift range. The power-law index can
be derived using a multiple regression method. Second, marginalize
the coefficient value over a reasonable range.

However, the gravitational lensing by random fluctuations in the
intervening matter distribution induces a dispersion in GRB
brightness \citep{Oguri06,Schaefer07}, degrading their value as
standard candles as well as SNe Ia \citep{Holz98}. GRBs can be
magnified (or reduced) by the gravitational lensing produced by the
structure of the Universe. The gravitational lensing has sometimes a
great impact on high-redshift GRBs. First, the probability
distribution functions (PDFs) of gravitational lensing magnification
have much higher dispersions and are markedly different from the
Gaussian distribution \citep{Valageas00,Oguri06,Wangfy11}. Figure \ref{magnification}
shows the magnification probability distribution functions of
gravitational lensing at different redshifts \citep{Wangfy11}.
Second, there is effectively a threshold for the detection in the
burst apparent brightness. With gravitational lensing, bursts just
below this threshold might be magnified in brightness and detected,
whereas bursts just beyond this threshold might be reduced in
brightness and excluded. \cite{Wangfy11} considered the weak lensing
effect on cosmological parameters derived from GRBs, and found that
the most probable value of the observed matter density $\Omega_M$ is
slightly lower than its actual value, see Figure \ref{CDM}. The weak
gravitational lensing also affects the dark energy equation of state
by shifting it to a more negative value.

\begin{figure}
\includegraphics[width=0.5\textwidth]{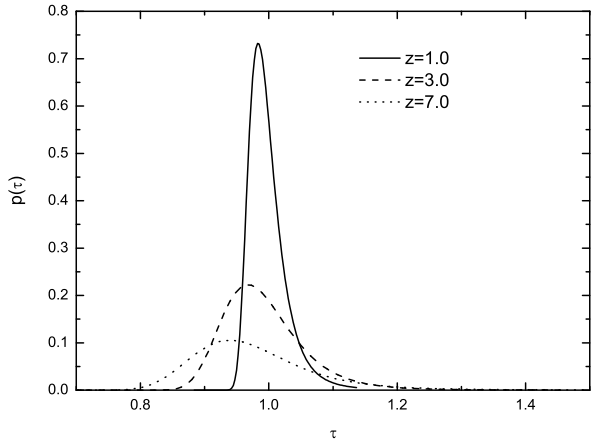}
\caption{Magnification probability distribution functions of
gravitational lensing at redshifts $z=1$, $z=3$ and $z=7$. (Adapted
from Figure 5 in \cite{Wangfy11}.)
 \label{magnification}}
\end{figure}

\begin{figure}
\includegraphics[width=0.5\textwidth]{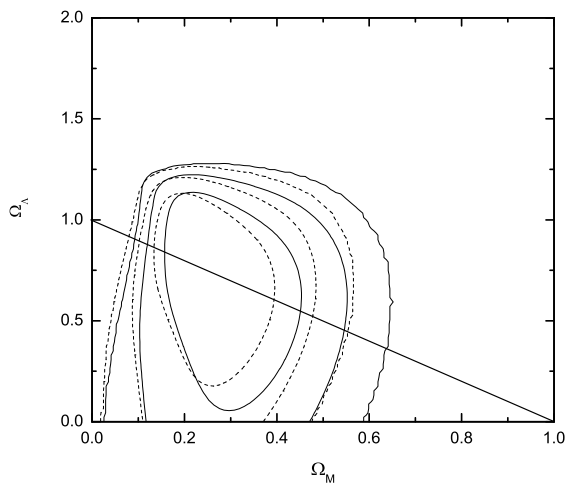}
\caption{Confidence contours of likelihood from $1\sigma$ to
$3\sigma$ in the $\Lambda$CDM model. The black line contours from
116 GRBs and the dotted contours from 116 GRBs including
magnification bias. (Adapted from Figure 7 in \cite{Wangfy11}.)
 \label{CDM}}
\end{figure}

\subsection{The equation of state of dark energy}
The dark energy equation of state $w$ is the most important parameter
that describes the properties of dark energy. Whether and how it
evolves with time is crucial for revealing the physics of dark
energy. GRBs can provide the high-redshift evolution property of
dark energy. The procedure is to bin $w$ in $z$,
and fit the $w$ in each bin to observational data by assuming that
$w$ is constant in each bin. The function $f(z)$ in equation
(\ref{eq:fz}) should be described as
\begin{equation}
  \label{eq:fzbinned}
  f(z_{n-1}<z \le z_n)=
  (1+z)^{3(1+w_n)}\prod_{i=0}^{n-1}(1+z_i)^{3(w_i-w_{i+1})},
\end{equation}
where $w_i$ is the EOS parameter in the $i^{\mathrm{th}}$ redshift
bin defined by an upper boundary at $z_i$, and the zeroth bin is
defined as $z_0=0$. \cite{Qi08a} used GRBs and other cosmological
observations to construct evolution of the equation of state, and
found that the equation of state $w$ is consistent with the cosmological
constant \citep[also see][]{Qi08b}. The confidence interval of the
uncorrelated equation of state parameter can be significantly
reduced by adding GRBs. After calibrating the GRB correlations using
cosmographic parameters, \cite{Wangfy11} found that the
high-redshift ($1.4<z<8.2$) equation of state is consistent with the
cosmological constant. But some studies found that the equation of
state $w$ may deviate from $-1$ \citep[i.e.,][]{Qi09,Zhao12}. In
light of the Planck CMB data, \cite{Wang14} found that the EOS is
consistent with the cosmological constant at the 2$\sigma$
confidence level, not preferring to a dynamical dark energy, which is
shown in Figure \ref{Wangf14}.

\begin{figure}
\includegraphics[width=0.5\textwidth]{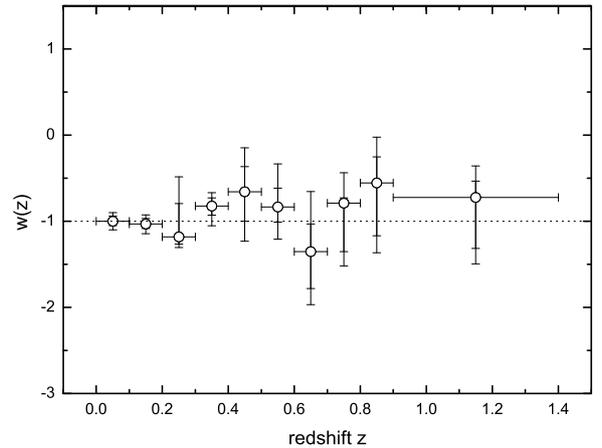}
\caption{Estimation of the uncorrelated dark energy EOS parameters
at different redshift bins ($w_1,w_2,...,w_{10}$) from SNe
Ia+BAO+WMAP9+H(z)+GRB data. The open points show the best fit value.
The error bars are $1\sigma$ and $2\sigma$ confidence levels. The
dotted line shows the cosmological constant. (Adapted from Figure 3
in \cite{Wang14}.) \label{Wangf14}}
\end{figure}

\section{Probing the star formation rate}

\subsection{Star formation rate derived from GRBs}\label{sec:SFR}

The association of long GRBs with core-collapse supernovae has been
confirmed from observations in recent years
\citep{Stanek03,Hjorth03}, which provides a complementary technique
for measuring the high-redshift SFR
\citep{Totani97,Wijers98,Lamb00,Porciani01,Bromm02}. The selection
effects should be considered \citep[for a review, see][]{Coward07}.
But one crucial problem appears, i.e., how to calibrate the GRB
event rate to the SFR. The luminosity function may play an important
role
\citep{Natarajan05,Daigne06,Salvaterra07,Salvaterra09,Campisi10,
Wanderman10,Cao11}. Before the launch of \emph{Swift}
\citep{Gehrels04}, the luminosity function is determined by fitting
the observed $\log N - \log P$ distribution
\citep{Schmidt99,Porciani01,Guetta05,Natarajan05}. Thanks to the
\emph{Swift}, more redshifts of GRBs are measured. This makes it
possible to give more information on the luminosity function
\citep{Wanderman10,Cao11,Tan13}. Because the form of luminosity
function should be assumed and the model parameters of luminosity
function is degenerate, it is not easy to determine the luminosity
function. A straightforward way to estimate the luminosity function
is proposed by \cite{LyndenBell71} and then further developed by
\cite{Efron92}. This method has been used for GRBs
\citep{LlydRonning02,Yonetoku04,Wu12}. There are two luminosity
function models in the literature, a broken power law
and a single power law with an exponential cut-off at low
luminosities.

\begin{figure}
\includegraphics[width=0.5\textwidth]{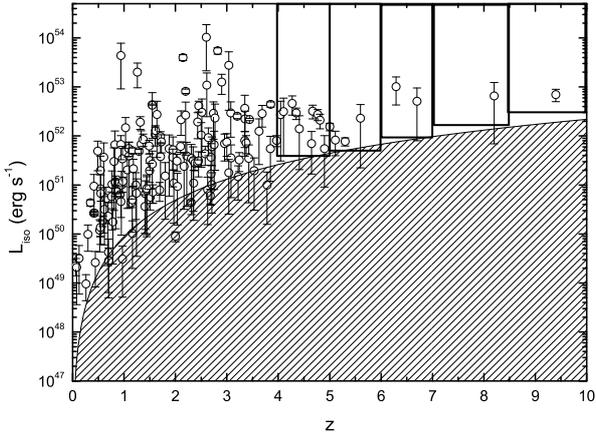} \caption{Distribution of the
isotropic-equivalent luminosity for 157 long-duration \emph{Swift}
GRBs. The shaded area approximates the detection threshold of
\emph{Swift} BAT. (Adapted from Figure 1 in \cite{Wang13}.)}
\label{GRBnum}
\end{figure}

In order to avoid the poorly known luminosity function when studying
high-redshift SFR, a method that only high-luminosity GRBs are used
is proposed \citep{Yuksel08,Kistler09,Wang091,Yu12,Wang13}. The
expected redshift distribution of GRBs is
\begin{equation}
\frac{d N}{d z}=F(z) \frac{\varepsilon(z)\dot{\rho}_*(z)}{\langle
f_{\rm beam}\rangle} \frac{dV_{\rm com}/dz}{1+z},
\end{equation}
where $F(z)$ represents the ability to obtain the redshift,
$\varepsilon(z)$ accounts for the fraction of stars producing GRBs, and
$\dot{\rho}_*(z)$ is the SFR density. The $F(z)$ can be treated as
constant when we consider the bright bursts with luminosities
sufficient to be detected within an entire redshift range. GRBs that
are unobservable due to beaming are accounted for through $\langle
f_{\rm beam}\rangle$. The $\varepsilon(z)$ can be parameterized as
$\varepsilon(z)=\varepsilon_0(1+z)^\delta$, where $\varepsilon_0$ is
an unknown constant that includes the absolute conversion from the
SFR to the GRB rate in a given GRB luminosity range.
\cite{Kistler08} found the index $\delta=1.5$ from 63 \emph{Swift}
GRBs. A little smaller value $\delta \sim 0.5-1.2$ has been inferred
from update \emph{Swift} GRBs \citep{Kistler09,Wang13}. In a flat
universe, the comoving volume is calculated by
\begin{eqnarray}
    \frac{dV_{\rm com}}{d z} = 4\pi D_{\rm com}^2 \frac{dD_{\rm com}}{d z} \;,
\end{eqnarray}
where the comoving distance is
\begin{eqnarray}\label{com}
    D_{\rm com}(z) \equiv \frac{c}{H_0} \int_0^z \frac{dz^\prime}{\sqrt{
            \Omega_m(1+z^\prime)^3 + \Omega_\Lambda}} \;.
\end{eqnarray}
In the calculations, the $\Lambda$CDM model with $\Omega_m=0.27$,
$\Omega_\Lambda=0.73$ and $H_0$=71 km~s$^{-1}$~Mpc$^{-1}$ from the
\emph{Wilkinson Microwave Anisotropy Probe} (WMAP) seven-year data
is used \citep{Komatsu11}.

Figure \ref{GRBnum} shows the isotropic luminosity distribution of
157 \emph{Swift} GRBs. The isotropic luminosity can be obtained by
\begin{equation}
L_{\rm iso}=E_{\rm iso}(1+z)/T_{90},
\end{equation}
where $T_{90}$ is the duration time. The shaded area approximates
the detection threshold of \emph{Swift} BAT, which has a flux limit
$\sim F_{\lim} = 1.2 \times 10^{-8}$erg cm$^{-2}$ s$^{-1}$. So the
selection effect is important. In order to exclude faint low
-redshift GRBs that could not be visible at high redshifts, we only
select luminous bursts. The luminosity cut $L_{\rm iso}> 10^{51}$
erg s$^{-1}$ is chosen in the redshift bin $0-4$ \citep{Yuksel08},
which removes many low-redshift, low-luminosity bursts that could
not be detected at higher redshift. The cumulative distribution of
GRB redshift can be expressed as
\begin{equation}
\frac{N(<z)}{N(<z_{\rm max})}=\frac{N(0,z)}{N(0,z_{\rm max})}.
\end{equation}
The value of $z_{\rm max}$ is taken as 4.0. Because the SFR has been
well measured at $z<4.0$ \citep{Hopkins06}. The theory predicted and
observed cumulative GRB distributions is shown in Figure~\ref{cum}.
The Kolmogorov-Smirnov statistic gives the minimization for
$\delta=0.5$ \citep{Wang13}. At the $2\sigma$ confidence level, the
value of $\delta$ is in the range $-0.15<\delta<1.6$.

\begin{figure}
\includegraphics[width=0.5\textwidth]{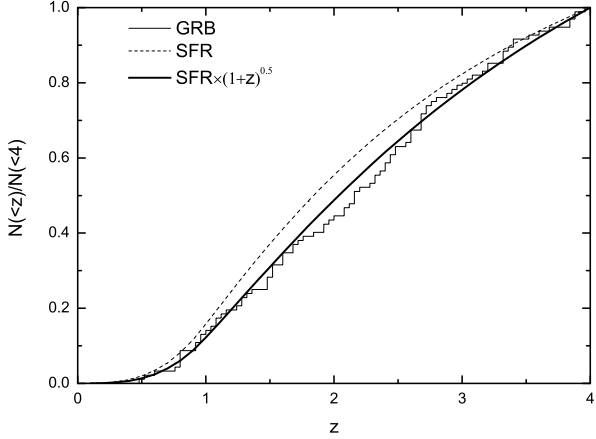} \caption{Cumulative distribution of 92 \emph{Swift} GRBs with
$L_{\rm iso}>10^{51}\rm erg~s^{-1}$ in $z=0-4$ (stepwise solid
line). The dashed line shows the GRB rate inferred from the star
formation history of Hopkins \& Beacom (2006). The solid line shows
the GRB rate inferred from the star formation history including
$(1+z)^{0.5}$ evolution. (Adapted from Figure 2 in \cite{Wang13}.)}
\label{cum}
\end{figure}

There are four redshift bins, $z=4-5$, $5-6$, $6-7$, $7-8.5$ and
$8.5-10$. The GRBs in $z = 1-4$ play as a ``control group'' to
constrain the GRB-to-SFR conversion. The theoretically predicated
number of GRBs in this bin can be calculated as
\begin{eqnarray}
N_{1-4}^{\rm th} & = & \Delta t \frac{\Delta \Omega}{4\pi}
\int_{1}^{4} dz\,  F(z) \, \varepsilon(z) \frac{\dot{\rho} _*(z)}
{\left\langle f_{\rm beam}\right\rangle} \frac{dV_{\rm com}/dz}{1+z} \nonumber \\
& = & A \, \int_{1}^{4} dz\, (1+z)^\delta \, \dot{\rho} _*(z) \,
\frac{dV_{\rm com}/dz}{1+z}\,, \label{N1-4}
\end{eqnarray}
where $A = {\Delta t \, \Delta \Omega \, F_0} / 4\pi {\left\langle
f_{\rm beam} \right\rangle}$ depends on the total observed time of
\emph{Swift}, $\Delta t$, and the angular sky coverage, $\Delta
\Omega$. The theoretical number of GRBs in redshift bin $z_1-z_2$ is
\begin{eqnarray}
N_{z_1-z_2}^{\rm th} & = &  \left\langle \dot{\rho} _*
\right\rangle_{z_1-z_2} A \, \int_{z_1}^{z_2} dz\, (1+z)^\delta \,
\frac{dV_{\rm com}/dz}{1+z}\,, \label{Nz1-z2}
\end{eqnarray}
where $\left\langle \dot{\rho}_* \right\rangle_{z_1-z_2}$ is the SFR
in the redshift range $z_1-z_2$. Representing the predicated
numbers, $N_{z_1-z_2}^{\rm th}$ with the observed GRB counts,
$N_{z_1-z_2}^{\rm obs}$, we obtain the SFR in the redshift range
$z_1-z_2$,
\begin{equation}
\left\langle \dot{\rho}_* \right\rangle_{z_1-z_2} =
\frac{N_{z_1-z_2}^{\rm obs}}{N_{1-4}^{\rm obs}} \frac{\int_{1}^{4}
dz\, \frac{dV_{\rm com}/dz}{1+z}(1+z)^\delta \dot{\rho}_*(z)\,
}{\int_{z_1}^{z_2} dz\, \frac{dV_{\rm com}/dz}{1+z}(1+z)^\delta }\,.
\label{zratio}
\end{equation}
The derived SFR from GRBs are shown as filled circles in
Figure~\ref{SFR}. Error bars correspond to 68\% Poisson confidence
intervals for the binned events \citep{Gehrels86}. The high-redshift
SFRs obviously decrease with increasing redshifts, although an
oscillation may exist. The SFRs from GRBs are dramatically larger
than those from other observations. The main reason is that other
observations probe only the brightest galaxies, especially at high
redshifts. But GRBs can reveal the faint galaxies at high redshifts
due to their high luminosity. The SFR at $z>4.48$ is proportional to
$(1+z)^{-3}$, which is shown as solid line in Figure~\ref{SFR}.

\begin{figure}
\includegraphics[width=0.5\textwidth]{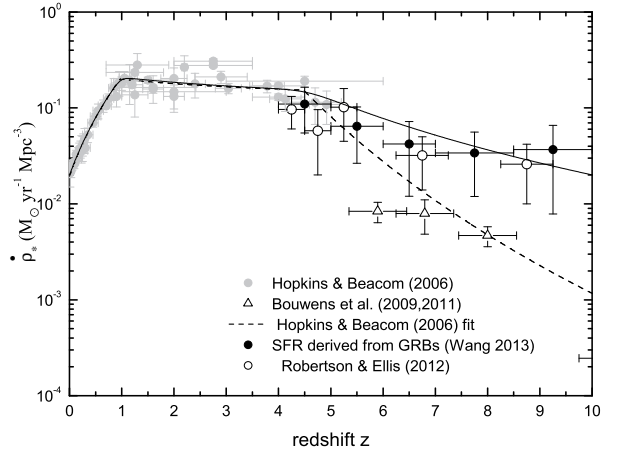} \caption{The cosmic star formation history.
The grey points are taken from Hopkins \& Beacom (2006), the dashed
line shows their fitting result. The triangular points are from
Bouwens et al. (2009, 2011). The open circles are taken from
\cite{Robertson12}. The filled circles are the SFR derived from GRBs
in \cite{Wang13}. (Adapted from Figure 3 in \cite{Wang13}.)}
\label{SFR}
\end{figure}

\subsection{Possible origins of high-redshift GRB rate excess}

Recent studies show that the rate of GRBs does not strictly follow
the SFH but is actually enhanced by some mechanism at high redshift
\citep{Le07,Salvaterra07,Kistler08,Yuksel08,Wang09,Robertson12,Wang13}.
The SFR inferred from the high-redshift ($z>6$) GRBs seems to be too
high in comparison with the SFR obtained from some high-redshift
galaxy surveys \citep{Bouwens09,Bouwens11}.

\subsubsection{Metallicity evolution}

A natural origin of the high-redshift GRB rate excess is the
metallcity evolution. Theory and observation both support that long
GRBs prefer to occurring in low-metallicity environment. Some
theoretical studies of long GRBs progenitors using stellar evolution
models suggest that low metallicity may be a necessary condition for
a long GRB to occur. For popular collapse models of long GRBs, stars
with masses $>30M_\odot$ can be able to create a black hole (BH)
remnant \citep{Woosley93,Hirschi05}. The preservation of high
angular momentum and high-stellar mass at the time of collapse
\citep{Woosley93,MacFadyen99} is crucial for producing a
relativistic jet and high luminosity. Low-metallicity
($0.1-0.3Z_\odot$) progenitors can theoretically retain more of
their mass due to smaller line-driven stellar winds
\citep{Kudritzki00,Vink05}, and hence preserve their angular
momentum \citep{Yoon05,Yoon06,Woosley061}, because the wind-driven
mass loss of massive stars is proportional to the metallicity.
Observations of long GRB host galaxies also show that they are
typically in low metallicity environment, for several local long GRB
host galaxies \citep{Sollerman05,Stanek06}, as well as in distant
long GRB hosts \citep[i.e.,][]{Fruchter06,Prochaska07}.

\cite{Li08} studied the possibility of interpreting the observed
discrepancy between the GRB rate history and the star formation rate
history using cosmic metallicity evolution \citep{Kistler08}. Under
the assumption that the formation of long GRBs follows the cosmic
star formation history and form preferentially in low-metallicity
galaxies, the rate of GRB is given by
\begin{equation}
R_{\rm GRB}(z)=k_{\rm GRB}\Sigma (Z_{\rm th},z)\rho _*(z),
\end{equation}
where $k_{\rm GRB}$ is the GRB formation efficiency, $\Sigma (Z_{\rm
th},z)$ is the fraction of galaxies at redshift $z$ with metallicity
below $Z_{\rm th}$ \citep{Langer06} and $\rho _*(z)$ is the observed
SFR. The function $\Sigma (Z_{\rm th},z)$ is \citep{Langer06}
\begin{equation}
\Sigma (Z_{\rm th},z)=\frac{\hat{\Gamma}[\alpha_1+2,(Z_{\rm
th}/Z_{\odot})^2 10^{0.15\beta z}]}{\Gamma(\alpha_1+2)},
\end{equation}
where $\hat{\Gamma}$ and $\Gamma$ are the incomplete and complete
gamma functions, $\alpha_1=-1.16$ and $\beta=2$ \citep{Savaglio05}.
\cite{Li08} found that the distribution of luminosity and cumulative
distribution of redshift could be well fitted if $Z_{\rm
th}=0.3Z_{\odot}$ is adopted. \cite{Wang091} studied the
high-redshift SFR by considering the GRBs tracing the star formation
history and the cosmic metallicity evolution. They found the SFR
derived from GRBs is marginal consistent with that from traditional
way \citep[i.e.,][]{Hopkins06}. Using Monte Carlo simulations,
\cite{Qin10} compared the simulation results to the Swift
observations with $\log N-\log P$ and luminosity-redshift
distributions. They found that the observed distributions are well
consistent with that from simulations if the GRB rate is
proportional to the SFR incorporating with the cosmic metallicity
history with $Z_{\rm th}=0.6Z_{\odot}$. Figure \ref{mc} shows the
comparison between simulations and observation. \cite{Wei14}
examined the influence on the GRB distribution due to the background
cosmology, i.e., $R_h=ct$ Universe. However, a few GRB hosts with
high metallicity are observed (i.e. GRB 020819), so that the role of
metallicity in driving the GRB phenomena remains unclear and it is
still debated \citep{Price07,Wolf07,Kocevski09,Graham09,Svensson10}.
For excellent reviews, see \cite{Fynbo12} and \cite{Levesque14}. But
there are some uncertainties when measure the metallicities of GRBs'
explosion region at high-redshifts, such as chemical inhomogeneity
\citep{Levesque10,Niino11}. \cite{WangD14} studied the metallicity
role from two aspects, the GRB host galaxies and redshift
distribution. They found that the the observed GRB host galaxy
masses and the cumulative redshift distribution can fit the
predicted distributions well if GRBs occur in low-metallicity $12 +
\log \rm(O/H)_{\rm KK04} < 8.7$, which is shown in Figure
\ref{host}. Trenti et al. (2015) found that there is clear evidence
for a relation between SFR and GRB \citep{Jimenez13}. But a sharp
cut-off of metallicity is ruled out.

\begin{figure*}
\includegraphics[width = 0.45 \textwidth]{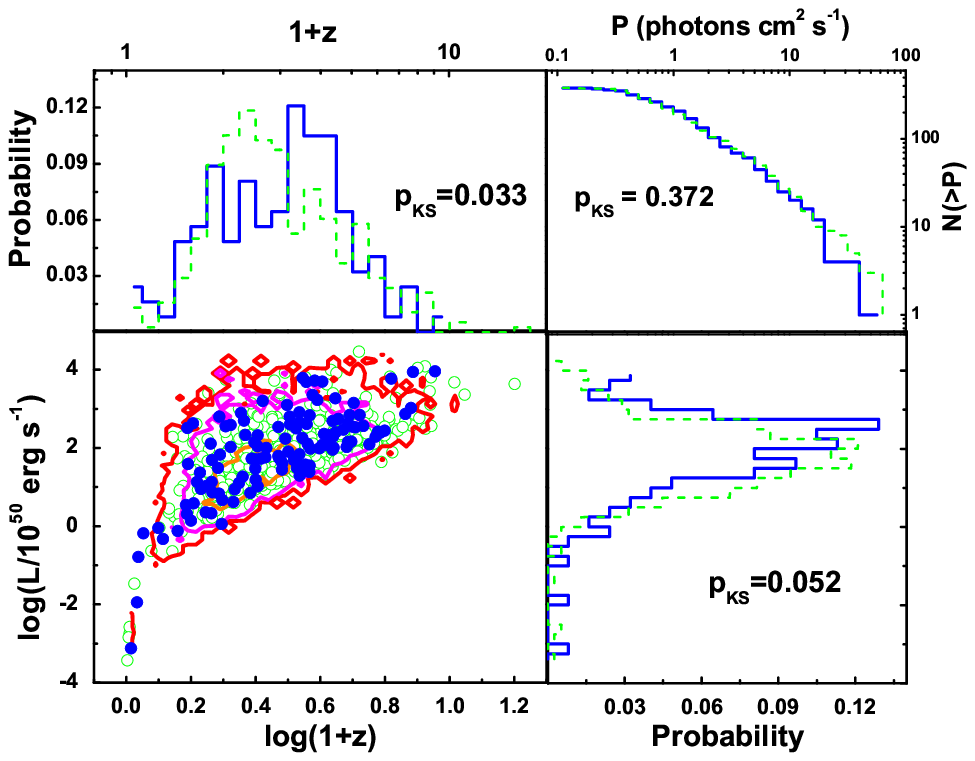}
\includegraphics[width = 0.45 \textwidth]{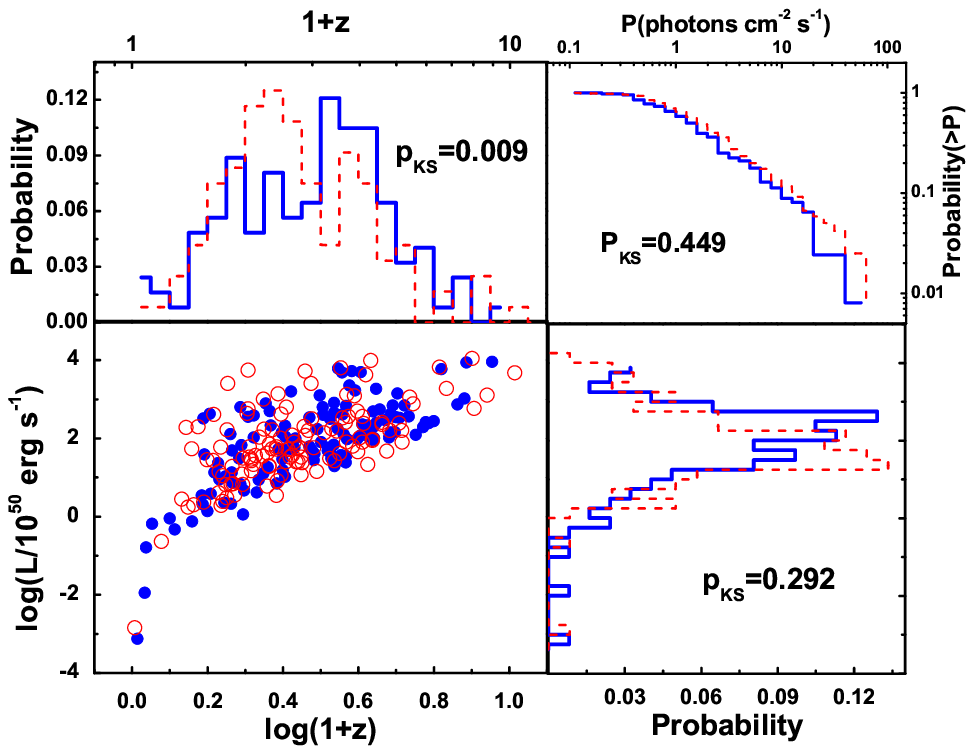}
\caption{Comparisons of $\log L-\log z$ distributions and $\log L$,
$\log z$, and $\log P$ distributions between the observed {\em
Swift}/BAT GRB sample (solid) and simulations (open dots and dashed
lines) for: $R_{\rm GRB}(z) \propto \rho _*(z)\times \Sigma (Z_{\rm
th},z)$. {\em left four panels} are for the trigged GRB sample. {\em
Right four panels} are for the sample with redshift measurement. One
dimensional K-S test probabilities for the comparisons are presented
in each panel. (Adapted from Figure 5 in \cite{Qin10}.)} \label{mc}
\end{figure*}

\begin{figure}
\includegraphics[width=0.5\textwidth]{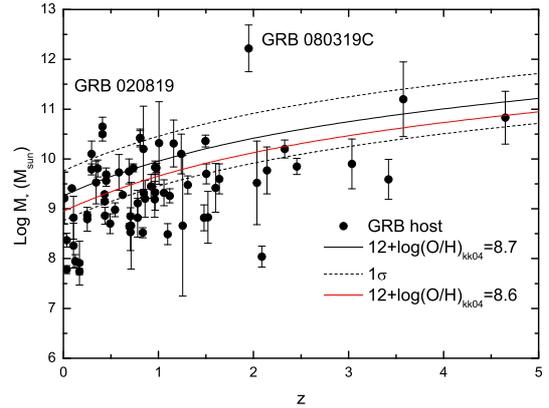} \caption{GRB host galaxy mass distribution.
The solid lines represent the upper limits of the stellar mass of a
GRB host galaxy given a metallicity cutoff of $12+\log\rm(O/H)_{\rm
KK04}=8.7$ (black), and $12+\log\rm(O/H)_{\rm KK04}=8.6$ (red). The
dashed lines represent the 1$\sigma$ scatter. (Adapted from Figure 4
in \cite{WangD14}.)} \label{host}
\end{figure}

\subsubsection{Evolving star initial mass function}
\cite{Wangf11} proposed that the GRB rate excess may be due to the
evolution of star initial mass function (IMF), also see
\citep{Xu08}. Because an ``top-heavy" IMF will lead to more massive
stars at high-redshift which can result in much more GRBs.
Considering long GRBs trace SFR, the rate of GRBs in an evolving IMF
is
\begin{equation}
R_{\rm GRB}  \propto \frac{{N_{m > 30M_ \odot  } }}{V} =K\left(
{\frac{c}{{H_0 }}} \right)^{ - 3} \frac{{\int_{30M_ \odot  }^{m_l }
{\xi (m)d\log m} }}{{\int_{m_s }^{m_l } {m\xi (m)d\log m} }}\rho
_*(z),
\end{equation}
where $K$ is a constant to be constrained and $R_{\rm GRB}$ is the
rate of GRBs, representing the number of GRBs per unit time per unit
volume at redshift $z$. The evolving IMF proposed by \cite{Dave08}
is
\begin{equation}
\frac{dN}{d\log{m}}=\xi(m)\propto\left\{
\begin{array}{l}
m^{-0.3}\;\; {\rm for}\; m<\hat{m}_{\rm IMF}\\
m^{-1.3}\;\; {\rm for}\; m>\hat{m}_{\rm IMF},
\end{array} \right.
\end{equation}
where $\hat{m}_{\rm IMF}=0.5 (1+z)^{2} M_\odot$, which has been
constrained by requiring non-evolving star formation activity
parameter. Figure \ref{cd} shows that the observed cumulative
distribution of GRBs can be well produced by this model.

\begin{figure}
\includegraphics[width = 0.45 \textwidth]{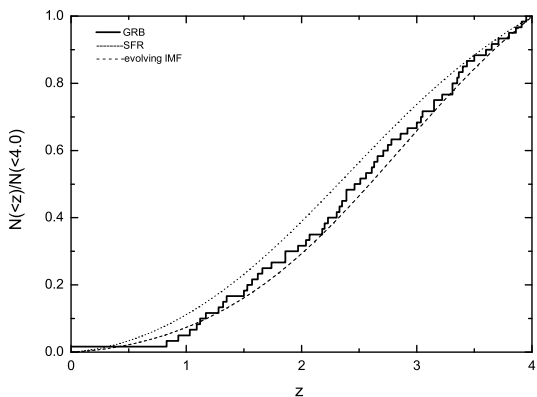} \caption{The cumulative distribution of
72 \emph{Swift} long GRBs with $L_{\rm iso}>0.8\times10^{51}$ erg
s$^{-1}$ (stepwise solid line). The dotted line shows the GRB rate
inferred from the star formation history of Hopkins \& Beacom
(2006). The dashed line shows the GRB rate inferred from star
formation history including an evolving IMF. (Adapted from Figure
2in \cite{Wangf11}.)}\label{cd}
\end{figure}

\subsubsection{Evolving luminosity function break}

\cite{Virgili11} found that if the break of luminosity function
evolves with redshift, the distributions of luminosity, redshift and
peak photon flux from the BATSE and Swift data can be reproduced
from simulations. The break luminosity function evolution can be in
a moderate way $\propto L_b\times (1+z)^{\sim 0.8-1.2}$.
\cite{Campisi10} studied the luminosity function, the rate of long
GRBs at high redshift, using high-resolution N-body simulations. A
strongly evolving luminosity function with no metallicity cut may
well explain the $\log N-\log P$ distribution of BATSE and Swift
data.

\subsubsection{Superconducting cosmic string}
Cosmic strings are thought to be linear topological defects that
could be formed at a phase transition in very early Universe. By
considering that high-redshift GRBs 080913 and 090423 are
electromagnetic bursts of superconducting cosmic strings,
\cite{Cheng10} showed the high-redshift GRB excess can be
reconciled. But \cite{WangY11} claimed that GRBs from cosmic string
have a very small angle, about $10^{-3}$, which could be in
contradiction with the opening angle of the GRB outflow.
\cite{Cheng11} pointed out that the angle is not the opening angle
of the GRB outflow, but is just the collimation angle of the
radiation of the corresponding string segment. We must caution that
the existence of cosmic string is only speculative.

\section{Probing the Pop III stars and High-Redshift IGM}

\subsection{Observational signature of Pop III GRBs}
The first stars, also called Population III (Pop III) stars, are
predicted to have formed in minihaloes with virial temperatures
$T_{\rm vir}\leq10^4$K at $z\geq15$ \citep{Tegmark97,Yoshida03,
Bromm04}. Numerical simulations show that Pop III stars forming in
primordial minihaloes, were predominantly very massive stars with
typical masses $M_*\geq100M_{\odot}$ \citep{Bromm99,Bromm02,Abel02},
for recent reviews, see \cite{Bromm09} and \cite{Bromm13}. They had
likely played a crucial role in early universe evolution, including
reionization, metal enrichment history. Some studies shows that some
Pop III stars will end as GRBs, called Pop III GRBs
\citep{Heger03,Bromm06,Komissarov10,Stacy11}, which will be brighter
and more energetic than any GRB yet detected
\citep{Toma11,Nagakura11,Campisi11,Meszaros10,Nakauchi12}. Direct
observations of the Pop III stars have so far been out of reach. The
properties of Pop III stars may be revealed by their remanents, Pop
III GRBs.

\begin{figure*}
\includegraphics[width = 0.45 \textwidth]{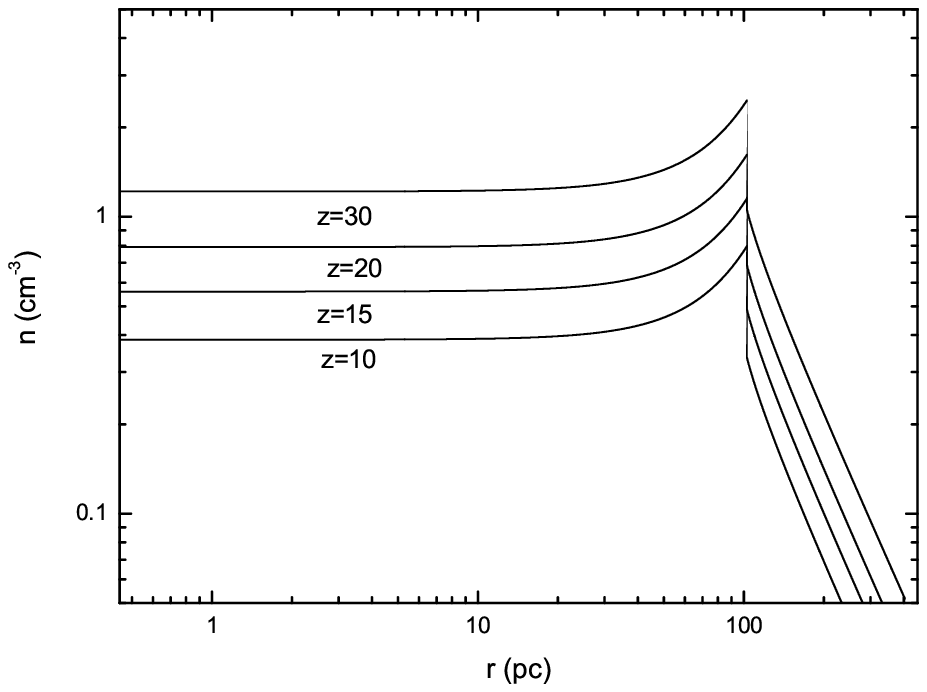}
\includegraphics[width = 0.45 \textwidth]{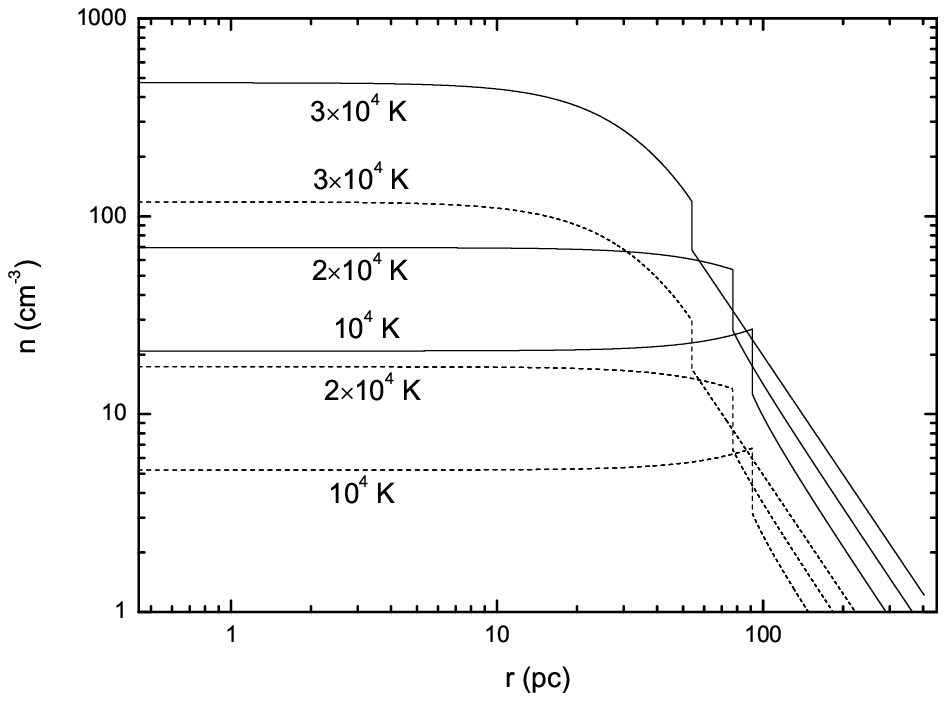} \caption{\textit{Left panel} Minihalo circumburst
density. Shown is the hydrogen number density as a function of
distance from the central Pop~III star at the moment of its death.
Typical circumburst densities are $\sim 1$\,cm$^{-3}$. \textit{Right
panel} Atomic-cooling-halo circumburst density. The density profiles
are calculated from the Shu solution. The case of photoheating from
only a single Pop~III star ({\it solid lines}), and that from a
stellar cluster ({\it dotted lines}).  (Adapted from Figures 1 and 2
in \cite{Wang12}.)} \label{Shu}
\end{figure*}

In order to predict the observational signature of Pop III GRBs, the
unusual circumburst environment that hosted the Pop III stars should
be determined. In particular, the properties of the afterglow
emission of Pop III GRBs depend on the circumburst density
\citep{Ciardi00,Gou04,Wang12}. In particular, the central minihalo
environments just before a massive star die and a GRB bursts out
can be understood as follows. The number of ionizing photons depends
strongly on the central stellar mass, which is determined by a
accretion flow onto the growing protostar
\citep[e.g.,][]{McKee08,Hosokawa11,Stacy12}. Meanwhile the accretion
is also affected by this radiation field. So the assembly of the Pop
III stars and the development of an H II region around them proceed
simultaneously, and affect each other. The shallow potential wells
of minihalos are unable to maintain photo-ionized gas, so that the
gas is effectively blown out of the minihalo. The resulting
photo-evaporation has been studied \citep{Alvarez06,Abel07,Greif09}.

The photoevaporation from minihalos can be described as the
self-similar solution for a champagne flow \citep{Shu02}. Assuming a
$\rho \propto r^{-2}$ density profile, the spherically symmetric
continuity and Euler equations for isothermal gas can be described
as follows:
\begin{equation} [ (v - x)^2 - 1]{1 \over \alpha} {d \alpha \over d
x}  =
 \left[\alpha - {2 \over x}(x - v) \right](x-v),
\label{alpha} \end{equation}
\begin{equation} [ (v - x)^2 - 1]{d v
\over d x} =
 \left[(x- v)\alpha - {2 \over x} \right](x-v),
\label{vel} \end{equation} where $x = r /c_s t$, and $\rho(r,t) =
\alpha(x)/4\pi G t^2=m_{\rm H} n(r)/X$ and $u(r,t)=c_s v(x)$ are the
reduced density and velocity, respectively. $c_s$ is the sound speed
and $X=0.75$ the hydrogen mass fraction. We set the typical lifetime
of a massive Pop~III star as $t=t_{\ast}\simeq 3 \times 10^6$\, yr.

In the left panel of Figure \ref{Shu}, we show the density profiles
at the end of the Pop~III progenitor's life in the minihalo case.
The circumstellar densities are nearly uniform at small radii. Such
a flat density profile is markedly different from that created by
stellar winds. But in the atomic cooling halo case, star formation and
radiative feedback is not well understood
\citep{Johnson09,SafranekShrader12}, such as the masses of stars,
and stellar multiplicity \citep{Clark11}. So we also use the
formalism of the Shu solution as above. We assume that either one
Pop~III star or a small stellar cluster forms. The densities are
shown in the right panel of Figure~\ref{Shu}. Similar to the
minihalo case, densities are nearly constant at small radii, but
overall values are much higher, which is due to the deep potential
wells, so that photoheated gas can easily be retained. Typical
circumburst densities are $n\sim 100$\,cm$^{-3}$. Pop~III GRBs
originating in atomic cooling halos may be extremely bright.

\begin{figure}
\includegraphics[width=0.5\textwidth]{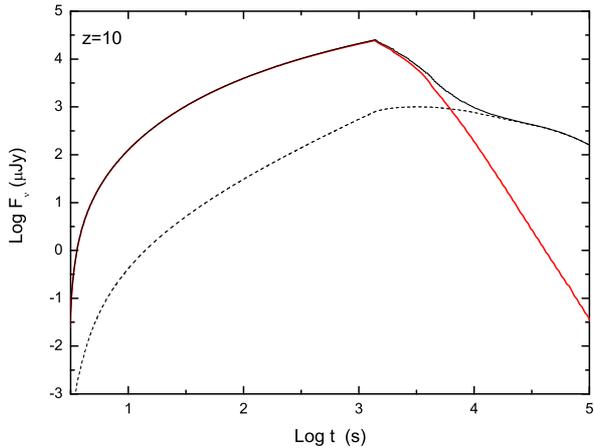} \caption{Light curve
at $\nu=6.3\times 10^{13}$Hz (M band) of Pop III GRBs. The emission
from the forward shock ({\it dashed line}), the reverse shock ({\it
solid red line}), and their combination ({\it solid black line}) are
shown. (Adapted from Figure 5 in \cite{Wang12}.) }\label{LCM}
\end{figure}

The typical parameters of the afterglow emission are adopted,
$\Gamma_0=300$, $E_{\rm iso}=10^{53}~\mbox{erg}$, $
\Delta_0=10^{12}~\mbox{cm}$, $\epsilon_e=0.3$, $\epsilon_B=0.1$, and
$p=2.5$. As an example, in Figure~\ref{LCM}, the M-band
($\nu=6.3\times 10^{13}$\,Hz) light curve is shown. Figure~\ref{LCK}
gives the observed flux at $\nu=1.36\times10^{14}$\,Hz as a function
of redshift in the minihalo case. The lines with filled dots, black
triangles and open dots correspond to an observed time of $6$
minutes, 1 hour, and 1 day respectively. The straight line marks the
K-band sensitivity for the near-infrared spectrograph (NIRSpec) on
James Webb Space Telescope \citep{Gardner06}. The high-redshift
cut-off is due to the Ly$\alpha$ absorption. The flux will be
completely absorbed by the intervening neutral IGM. At these
frequencies, the flux of afterglow is weakly dependent on redshift
of GRB. There are two reasons. First, the time dilation
effect implies that the high redshift means the earlier emission
times, where the afterglow are much brighter
\citep{Ciardi00,Bromm12}. Second, circumburst densities of GRBs
modestly increase with redshift.

\begin{figure}
\includegraphics[width=0.5\textwidth]{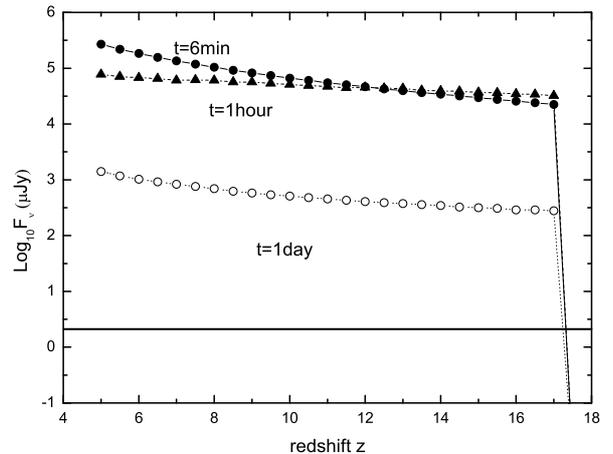} \caption{Observed flux at
$\nu=1.36\times 10^{14}$Hz (K band) as a function of redshift at
different observed times, as labelled. The K-band sensitivity of the
NIRSpec instrument on board the {\it JWST} is shown as a horizontal
line. The sharp cut-off at $z\simeq 17$ is due to Ly$\alpha$
absorption in the IGM. (Adapted from Figure 6 in
\cite{Wang12}.)}\label{LCK}
\end{figure}

\subsection{Metal enrichment history}
The metal enrichment history has several important influences for
cosmic structure formation. For example, the metal injection change
the mode of star formation ~\citep{Bromm01,Schneider02}. The
transition between Pop III star formation and ``normal" (Pop I/II)
star formation has important implications, e.g., the expected GRB
redshift distribution \citep{Bromm02}, reionization
\citep{Wyithe03}, and the chemical abundance patterns of stars. So
it is important to map the topology of pre-galactic metal
enrichment. Ten or thirty meter-class telescopes have been proposed
to measure the $z>6$ IGM metallicity with the GRB afterglow
\citep{Oh02}. Meanwhile, the relative gas column density from metal
absorption lines can reflect the enrichment history
\citep{Hartmann08,Wang12}.

Absorption processes and absorption lines imprinted on the spectra
of GRBs or quasars are the main sources of information about the
chemical and physical properties of high-redshift universe. But the
bright QSO number is very low at $z>6$ \citep{Fan06}. Meanwhile,
there are several high-redshift GRBs: GRB 050904 at $z=6.29$, GRB
080913 at $z=6.7$, GRB 090423 at $z=8.3$ and GRB 090429B at $z=9.4$.
The progenitors of long GRBs are thought to be massive stars, so the
number of high-redshift GRBs does not decrease significantly. The
density, temperature, kinematics and chemical abundances can be
extracted from absorption lines \citep{Oh02,Furlanetto03}. For
instance, \cite{Kawai06} have identified several metal absorption
lines in the afterglow spectrum of GRB 050904 and found that this
GRB occur in metal-enriched regions. Two absorption lines have been
observed in the spectrum of GRB 090423 at $z=8.2$
\citep{Salvaterra09}. These lines are due mainly to absorption metal
elements in low ionization stages.

\begin{figure*}
\includegraphics[width=0.5\textwidth]{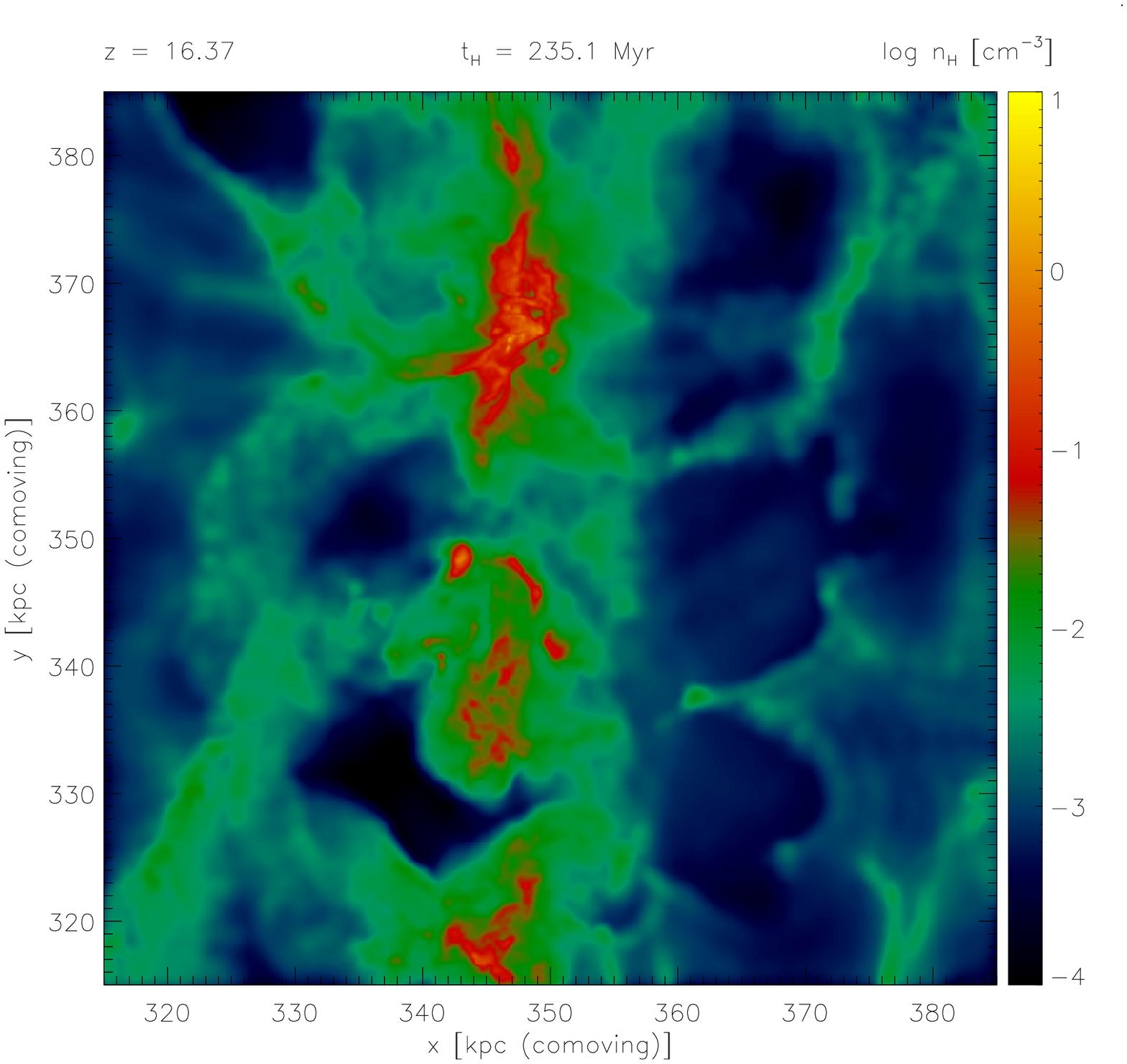}
\includegraphics[width=0.5\textwidth]{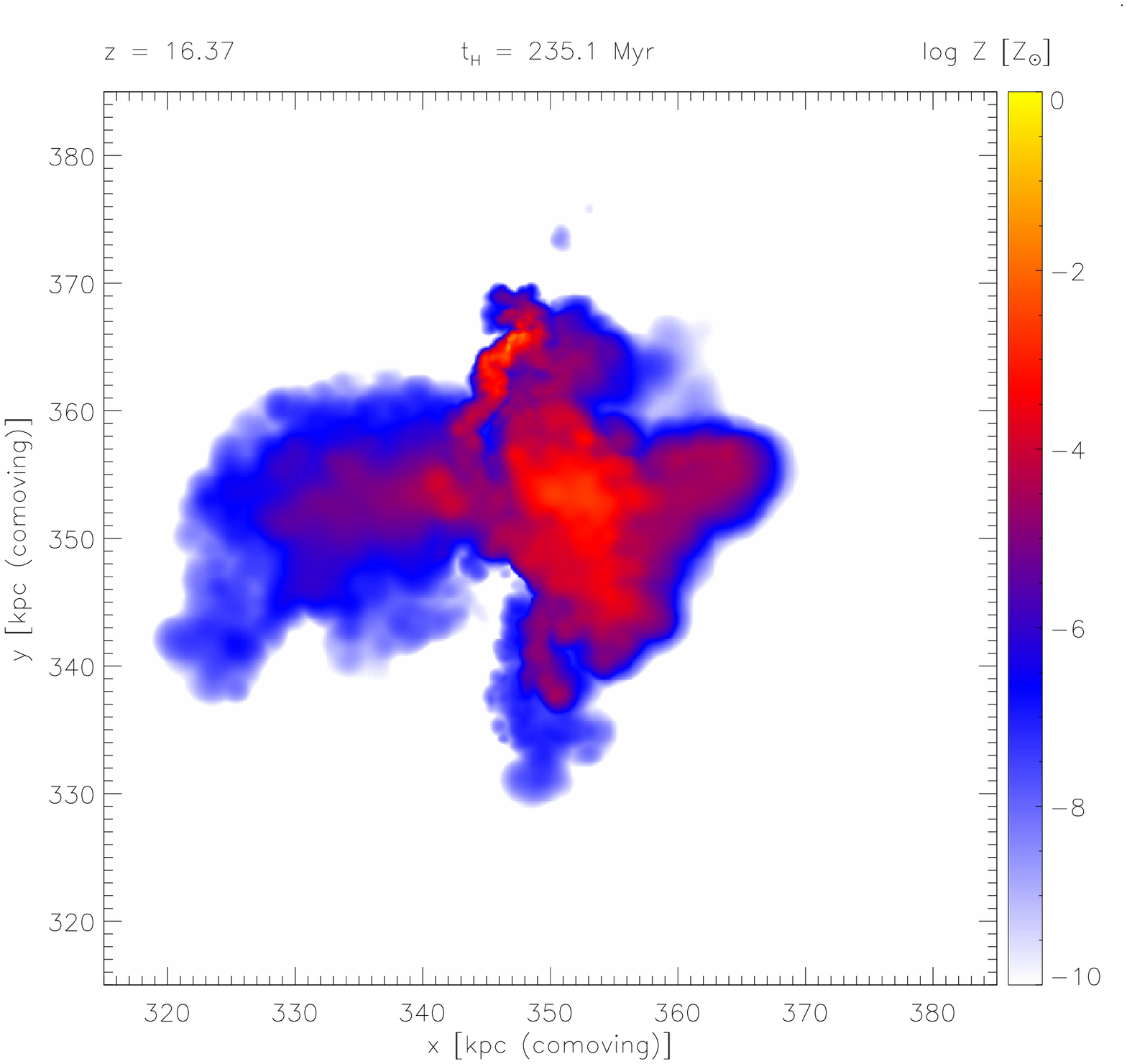}
\caption{Possible explosion sites for high-redshift GRBs. Shown are
the hydrogen number density (left panel) and metallicity contours
(right panel) averaged along the line of sight at $z\sim 16.37$,
when the first galaxy forms. The topology of metal enrichment is
highly inhomogeneous. (Adapted from Figure 3 in \cite{Wang12}.)}
\label{galsim}
\end{figure*}

\cite{Wang12} studied the ability of metal absorption lines in the
spectra of Pop III GRBs to probe the pre-galactic metal enrichment.
The first galaxy simulation carried out by \cite{Greif10} was used.
The simulation allowed one Pop~III progenitor star to explode as an
energetic supernova, then the IGM was polluted by the ejected
metals. The simulation box size is $ 1~\rm{Mpc}$ (comoving), and is
initialized at $z=99$ according to the $\Lambda$CDM model with
parameters: $\Omega_{\rm{m}}=1-\Omega_{\Lambda}=0.3$,
$\Omega_{\rm{b}}=0.04$,
$h=H_{0}/\left(100~\rm{km}~\rm{s}^{-1}~\rm{Mpc}^{-1}\right)=0.7$,
spectral index $n_{\rm{s}}=1.0$, and normalization $\sigma_{8}=0.9$
\citep{Spergel03}.

In Figure~\ref{galsim}, the hydrogen number density and metallicity
averaged along the line of sight are shown within the central
$\simeq 100$ kpc closer to the virialization of the first galaxy at
$z=16.4$. The distribution of metals produced by the first SN
explosion is highly inhomogeneous, and the metallicity can reach up
to $Z\sim 10^{-2.5} Z_{\odot}$, which is already larger than the
critical metallicity, $Z_{\rm crit}\leq 10^{-4} Z_{\odot}$.
Therefore, both Pop~III and Pop~I/II stars will form during the
assembly of the first galaxies \citep{Johnson08,Maio10}, so
simultaneous occurrence of Pop~III and normal GRBs at a given
redshift \citep{Bromm06,deSouza11}. We consider a Pop~III burst
exploding in one of the (still metal-free) first galaxy progenitor
minihalos at $z\simeq 16.4$.

For simplicity, we consider that prior to the GRB only one nearby SN
exploded beforehand, dispersing its heavy elements into the pristine
IGM. Two nucleosynthetic metal yields for Type~II core-collapse SNe
\citep{Woosley95}, and for pair-instability supernovae
\citep[PISNe;][]{Heger02,Heger10} are considered. Because the
hydrogen is substantially neutral, metals will reside in states
typical of C II, O I, Si II, and Fe II, because high-energy photons
able to further ionize these elements will be absorbed by H I
\citep{Furlanetto03}.

Figure~\ref{totalspeTC} shows two spectra of afterglow at the
reverse shock crossing time. Top panel is for the top-heavy (Very
Massive Star) initial mass function (PISN case) and bottom for
normal initial mass function (Type II SNe case). The cutoff is due
to Lyman-$\alpha$ absorption in the IGM which is expected to be
still completely neutral at $z>10$. In the two cases, the metal
lines are markedly different. The metal yields could be obtained
from metal lines. So the initial mass function of Pop III stars can
be derived from the metal absorption lines.

\begin{figure}
\includegraphics[width=0.5\textwidth]{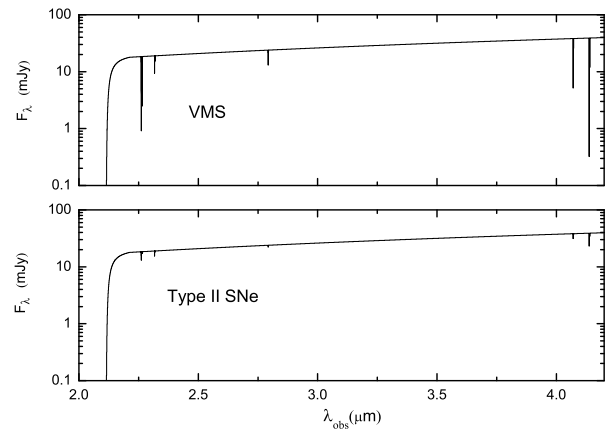} \caption{Pop III GRB spectrum observed
at the reverse shock crossing time
$t_{\oplus}=16.7\times(1+16.4)$\,s. Metal absorption lines are
imprinted according to the Pop~III SN event, PISN vs. core-collapse.
The former originates from a very massive star (VMS) progenitor,
whereas the latter from a less massive one. In each case, the cutoff
at short wavelengths is due to Lyman-$\alpha$ scattering in the
neutral IGM. Adapted from Figure 10 in \cite{Wang12}.)
}\label{totalspeTC}
\end{figure}

\begin{figure}
   \includegraphics[angle=-90,width=9cm]{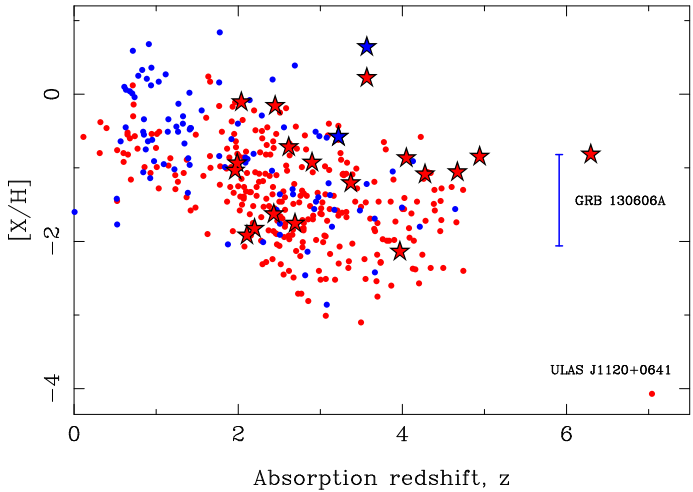}
\caption{The metallicity ([X/H]) as a function of redshift is shown
for QSO-DLAs (circles) and GRB-DLAs
\citep[stars][]{Schady11,Thone13}, including GRB 130606A at $z=5.91$
and ULAS J1120+0641 at z $\sim$ 7 \citep{Simcoe12}. Blue colors are
derived for log N(HI) $<$ 20.3 and red is derived for log N(H I)
$\geq$ 20.3. (Adapted from Figure 8 in \cite{CastroTirado14}.}
\label{fig:metal-z}
\end{figure}

Strong absorption lines detected in GRB spectra, called damped
Lyman-$\alpha$ (DLA) systems, could be used to probe the metal
rnrichment \citep{Savaglio06}. For example, \cite{CastroTirado14}
detected the DLA systems of GRB 130606A at $z=5.91$. The metallicity
of this GRB environment is in the range from $\sim 1/7$ to $1/60$ of
solar from the analysis of metal absorption lines. Figure
\ref{fig:metal-z} shows the metallicities derived from GRB-DLAs and
QSO-DLAs. The GRB130606A sub-DLA is only the third GRB absorber with
sub-DLA HI column density. So GRB sub-DLA is rare and hard to find.
\cite{Simcoe12} discovered the DLA of ULAS J1120+0641 at $z\sim 7$
with low metallicity. So GRB events at $z>10$ from future mission,
offer an exciting new window to probe pre-galactic metal enrichment
in high-redshift galaxies.

\subsection{Cosmic reionization}
The Gunn-Peterson (GP) test tells us that the IGM is almost fully
ionized at $z\leq5$ \citep{Gunn65}. Reionization of the IGM is
thought to have occurred during $z\sim6-20$ by Pop III stars and/or
quasars, and the precise measurement of the reionization is one of
the key topics in modern cosmology \citep[for reviews,
see][]{Barkana01,Robertson10}. The absorption of GRB afterglow is
dependent on the structure of reionization and the global history of
reionization, so it has the potential to distinguish between
different theoretical models of reionization
\citep{MiraldaEscude98,McQuinn08}. The afterglows at wavelengths
close to the Ly$\alpha$ resonance potentially provide a sensitive
probe of the ionization fraction in the IGM
\citep{MiraldaEscude98,Barkana04}. The IGM neutral fraction could be
derived by fitting red damping wing with high precision. The
absorption from host galaxy complicates the measurement of the IGM
ionization state from a GRB spectrum, but in principle this
absorption is less extended in wavelength and could be separated.
Some studies show that 20-30\% of GRB host galaxies have small HI
column density to allow determination of the absorption from a
partially ionized IGM \citep{Chen07}.

By studying the afterglow spectrum of GRB 050904 at $z=6.3$,
\cite{Totani06} found that the IGM was already largely ionized at
$z=6.3$, and the upper limit of $0.17$ for the neutral fraction of
IGM at 68\% confidence level. But the absorption from host galaxy
dominates the absorption redward of the Ly$\alpha$ forest, which
limits the constraints on reionization \citep{Totani06,McQuinn08}.
The bright optical afterglow of GRB 130606A at $z=5.9$ gives an
opportunity to probe the ionization status of IGM. The neutral
fraction of IGM is found to be 0.1 to 0.5 by analyzing of the red
Ly$\alpha$ damping wing of the afterglow spectrum taken by Subaru
\citep{Totani14}.

From theoretical view, the reionization process can also be studied
through theoretical model. The average evolution of $Q_{\rm
HII}=n_e/n_H$ is derived by numerical integration of the rate of
ionizing photons minus the rate of radiative recombinations
\citep{Madau99,Barkana01,Wyithe03,Yu12}
\begin{equation}
    \frac{dQ_{\rm HII}}{dz} = \left( \frac {\dot{N}_{\rm ion}} { n_H } -
           \alpha_B C n_H Q_{\rm HII}\right) \frac{dt}{dz}  \;  ,
\label{eq:ionization_history}
\end{equation}
where
\begin{equation}
\dot{N}_{\rm ion} =(1+z)^3 \dot{\rho}_*(z)N_\gamma f_{\rm esc}/m_p
\end{equation}
is the rate of ionizing photons ejected into the IGM, $N_\gamma$ is
the number of ionizing photons, $\dot{\rho}_*(z)$ is the SFR and
$f_{\rm esc}$ is the escape fraction. Using the SFR derived from
GRBs in section \ref{sec:SFR}, the evolution of the HII volume
filling factor $Q_{\rm HII}$ can be numerically calculated from
equation (\ref{eq:ionization_history}). Figure \ref{zre} shows the
evolution of $Q_{\rm HII}$ as a function of redshift.

\begin{figure}
\includegraphics[width=0.5\textwidth]{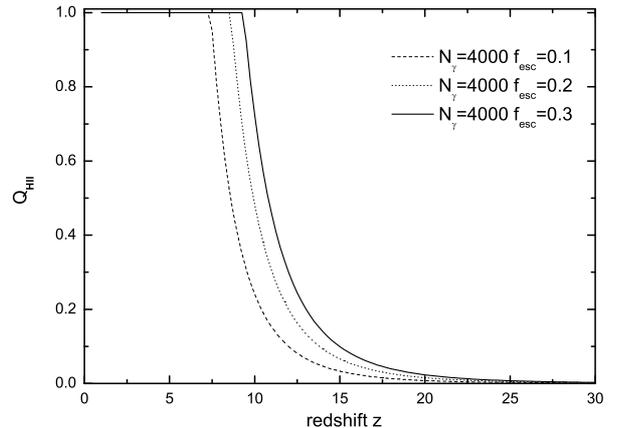} \caption{The HII filling factor $Q_{\rm HII}$ as a function of redshift computed for
different values of $f_{\rm esc}$. (Adapted from Figure 5 in
\cite{Wang13}.)} \label{zre}
\end{figure}

The cosmic microwave background (CMB) optical depth back to redshift
$z$ is also seriously depend on the reionization history, which can
be written as the integral of $n_e \sigma_T d \ell$, i.e.,
\begin{equation}
   \tau_e(z) = \int _{0}^{z} n_e(z) \sigma_T (1+z')^{-1} \; [c/H(z')] \; dz'    \;   .
\end{equation}
The optical depth is shown in Figure \ref{tau}. The WMAP nine-year
data gives $\tau_e=0.089\pm 0.014$ \citep{Hinshaw13}, which is shown
as the shaded region. The combination of Planck and WMAP data also
gives $\tau_e=0.089_{-0.014}^{+0.012}$ \citep{Planck13}. So
GRB-inferred SFR can reproduce the CMB optical depth. But the value
of the escape fraction $f_{\rm esc}$ \citep{Robertson10} and
clumping factor $C$ are hard to determined.

\begin{figure}
\centering
\includegraphics[width=0.5\textwidth]{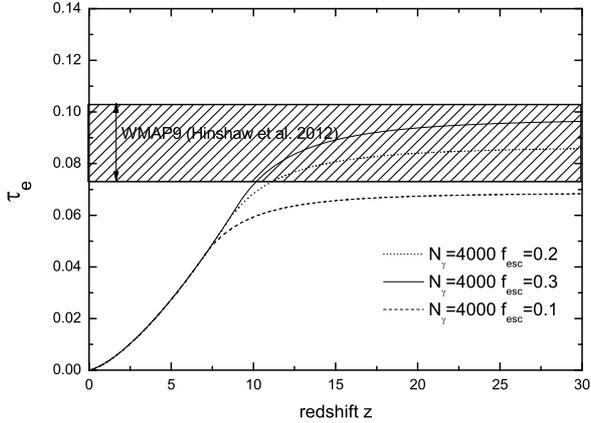} \caption{The optical depth $\tau_e$ due to the
scattering between the ionized gas and the CMB photons is shown. The
shade region is given by the nine-year WMAP measurements. The
reionization history calculated from GRB-inferred SFR can easily
reach $\tau_e$ from WMAP nine-year data. Adapted from Figure 6 in
\cite{Wang13}.)} \label{tau}
\end{figure}

\section{Summary and future prospect}
GRBs are observed throughout the whole electromagnetic spectrum,
from radio waves to $\gamma$-rays, which have been observed in
distant universe. Recently, GRBs have attracted a lot of attention
as promising standardizable candles to construct the Hubble diagram
to high redshift, as complementarity to other cosmological probes,
such as SNe Ia, CMB and BAO. However, a lot of work is needed to be
sure that GRBs can hold this promise in future. The most important
thing is to search for a correlation similar to that used to
standardize SNe Ia. In order to obtain the correlation, the
classification of GRBs may be crucial. We must remind that only SNe
Ia are standard candles among all SNe. The classical classification
method is based on the prompt emission properties (duration,
hardness, and spectral lag). The physics of prompt emission are not
fully understood \citep{Zhangb14}, and some new clues from other
objects are found \citep{Wangf13,Wangf14}. But observations of some
GRBs are challenging the standard classification
\citep{Zhang06,Zhang09,Lv10}. So more physical nature of GRBs is
needed \citep{Zhang09}. The circularity problem could be partially
solved by analyzing a sample of GRBs within a small redshift bin
\citep{Lamb05,Ghirlanda06,Liang06}. In particular, \cite{Liang06}
found that one can calibrate the power law indices of various
standard candle correlations with this method.

In order to measure high-redshift SFR from GRBs, the relation
between long GRB rate and SFR must be known. Besides, theoretical
models of the SFR have several free parameters, such as the
efficiency of star formation and the chemical feedback strength.
From the theoretical SFR, the predicted GRB redshift distribution
can be derived. So one can use the GRB redshift distribution
observed by Swift (or future missions such as SVOM and EXIST), to
calibrate the free parameters. More GRB red damping wing with low HI
column density are required to study properties of IGM.

Metal absorption lines in the GRB afterglow spectrum, giving rise to
EWs of a few tens of {\AA}, which may allow us to distinguish whether the
first heavy elements were produced in a Pop III star died as a
PISN or a core-collapse SN. To this extent, the spectrum needs to
be obtained sufficiently early, within the first few hours after the
trigger. Upcoming JWST would detect much more high-redshift GRBs
(properly Pop III GRBs) with high resolution NIR spectra including
metal absorption lines, which allow one to measure the cosmic
metallicity evolution.

In the future, the French-Chinese satellite Space-based multi-band
astronomical Variable Objects Monitor (SVOM) and JWST, have been
optimized to increase the number of GRB and the synergy with the
ground-based facilities. There are a combination of multi-wavelength
detectors on board of SVOM \citep{Paul11}. ECLAIRs wide-field camera
will detect GRBs in the energy range of 4-150 keV. The spectral
information of prompt emission will be measured by Gamma-Ray Monitor
(GRM). The afterglow can be obtained by the Micro channel X-ray
Telescope (MXT; 0.3-10 keV) and the Visible Telescope (VT;
400-900nm). SVOM can detect about 80 GRBs per year, and more than
50\% of GRBs have redshift measurement \citep{Petitjean11}. JWST is
a large, infrared-optimized space telescope with 6.6 m diameter
aperture. It has four scientific instruments: a Near-IR Camera
(NIRCam), a Near-IR Spectrograph (NIRSpec), a near-IR Tunable Filter
Imager (TFI), and a Mid-IR Instrument (MIRI) \citep{Gardner06}. But
the direct detection of a single Pop III star is not feasible even
for JWST, i.e., the AB magnitude of a $M=1000M_\odot$ star is only
36 at $z\sim 30$. Meanwhile, the Pop III GRBs can be detectable by
JWST \citep{Wang12,Mesler14,Macpherson13}. This will boost the
amount of information available to tackle the important issues
revealed by this exciting field of research.

\section*{Acknowledgements}
We thank the anonymous referee for detailed and very constructive
suggestions that have allowed us to improve our manuscript. We thank
Shuang-Nan Zhang for helpful discussions and comments. This work is
supported by the National Basic Research Program of China (973
Program, grant No. 2014CB845800), the National Natural Science
Foundation of China (grants 11422325, 11373022, and 11033002), the
Excellent Youth Foundation of Jiangsu Province (BK20140016), and the
Program for New Century Excellent Talents in University (grant No.
NCET-13-0279).

\end{document}